\documentclass[11pt,draft]{article}
\usepackage{latexsym}
\usepackage{amsmath}
\usepackage{amsthm}
\usepackage{amsfonts}
\allowdisplaybreaks
\title{Mixing Renormalization for Scalar Fields}
\author{Antonio O.\ Bouzas \thanks{E-mail:
    abouzas@mda.cinvestav.mx}\\\small Departamento de F\'{\i}sica
  Aplicada, CINVESTAV-IPN \\\small Carretera Antigua a Progreso Km.\
  6, Apdo.\ Postal 73 ``Cordemex''\\\small
  M\'erida 97310, Yucat\'an, M\'exico}
\date{December 9, 2002}
\newcommand{\sss}[1]{{\scriptscriptstyle #1}}
\newcommand{\ns}{\ensuremath{N_{s}}}
\newcommand{\nf}{\ensuremath{N_{f}}}
\newcommand{\ms}{\ensuremath{\mathrm{MS}}}
\newcommand{\msb}{\ms\ }
\newcommand{\os}{\ensuremath{\mathrm{OS}}}
\newcommand{\mat}[1]{\ensuremath{\boldsymbol{#1}}}
\newcommand{\vect}[1]{\ensuremath{\boldsymbol{\vec{#1}}}}
\newcommand{\Ll}{\ensuremath{\mathcal{L}}}
\newcommand{\V}{\ensuremath{\mathcal{V}}}
\newcommand{\sy}{\ensuremath{\mathcal{S}_\sss{Y}}}
\newcommand{\cy}{\ensuremath{\mathcal{C}_\sss{Y}}}
\newcommand{\sya}[1]{\ensuremath{\mathcal{S}_\sss{Y_{#1}}}}
\newcommand{\dy}{\ensuremath{d_\sss{Y}}}
\newcommand{\zz}{\ensuremath{\mathbb{Z}_{2}}}
\newcommand{\zy}{\ensuremath{\mathbb{Z}_{2\sss{Y}}}}
\newcommand{\zya}[1]{\ensuremath{\mathbb{Z}_{2\sss{Y_#1}}}}
\newcommand{\naught}{\sss{0}}
\newcommand{\dirac}{\not\negmedspace\partial}
\newcommand{\pirac}{\not\negmedspace p\,}

\newcommand{\tr}{\mathrm{Tr}}
\newcommand{\Tr}[1]{\mathrm{Tr}\left( #1 \right)}
\newcommand{\diag}{\mathrm{diag}}
\newcommand{\rank}{\mathrm{rank}}
\newcommand{\mub}{\overline{\mu}}
\newcommand{\pdot}{\!\cdot\!}
\renewcommand{\Re}{\ensuremath{\mathfrak{Re}}}

\newcommand{\xo}{{x^{\naught}}}
\renewcommand{\H}[1]{\overline{#1}}
\begin{document}
\maketitle
\begin{abstract}
  We consider the renormalization of theories with many scalar fields.
  We discuss at the one-loop level some simple, non-gauge models with
  an arbitrary number of scalars and fermions both in mass-shell and
  MS schemes.  In MS scheme we give a detailed qualitative analysis of
  the RG flow of dimensionless couplings in flavor space. 
\end{abstract}
\section{Introduction}
\label{sec:intro}

In this paper we consider the renormalization of theories with many
scalar fields which mix non-trivially under interactions.
Specifically, we discuss at the one-loop level two simple, non-gauge
models involving scalar fields and fermions.  We understand mixing of
fields in a broad sense, as the existence of flavor off-diagonal
propagators that do not vanish.  (The stronger condition that they
should not vanish on the mass shell is necessary for mixing of mass
eigenstates, but it will not be required in what follows.)  Mixing is
trivial if a flavor basis exists such that there is no mixing in that
basis.  We refer only to orthonormal flavor bases in which the kinetic
terms are diagonal and normalized, and to unitary or orthogonal
transformations among those bases.  Therefore, mixing is trivial if
the relevant two-point function can be diagonalized by a (coordinate
and momentum independent) unitary transformation, i.e., if it is
normal as a matrix in flavor space.  In the case of fermion fields
renormalizable, Lorentz-invariant interaction terms can only be
bilinear in the fields.  In the scalar sector a larger variety of
interactions is allowed and, as a consequence, mixing can be due to
interactions that are quadratic, cubic or quartic in the scalar
fields.

In many theories, like the Standard Model, there exists an
``interaction'' flavor basis which diagonalizes the bilinear
interaction terms.  More generally, we define an interaction basis as
one in which there would be no mixing if mass matrices were
proportional to the identity.  Such bases need not exist in general,
however, as would happen in a model with two different fermion
currents which are not simultaneously diagonalizable.  This phenomenon
is even more obvious in theories with mixing of scalar fields.  Due to
the presence of the different types of interactions mentioned above it
is often impossible to find a flavor basis in which there is no
mixing, even if mass terms are trivial.  Our treatment is independent
of the existence of an interaction basis.

In the next section we study a model with \ns\ scalar and \nf\ fermion
fields.  The scalar fields mix through a bilinear term involving a
mixing matrix (and an additional field taken also to be a scalar),
quartic self-interactions and Yukawa interactions.  The model does not
possess continuous symmetries and is intended to minimize the effect
on mixing of the quartic terms in the potential as much as possible.
We discuss the renormalization of this model in detail at the one-loop
level, for generic \ns\ and \nf, both in on-shell (\os) and \msb\ 
schemes.  In \msb\ we pay particular attention to the renormalization
group (RG) flow of dimesionless parameters.

The treatment of counterterms in section \ref{sec:1model} follows
closely our previous paper \cite{bou} on fermion mixing, with which
we try to keep the overlap at a minimum.  Besides the different nature
of the interaction Lagrangian for scalar and fermion fields, another
important difference with \cite{bou} is that in this paper we
consider the case of several coupling matrices, Yukawa couplings being
row matrices, whereas \cite{bou} is restricted to models with only
one. The presence of several coupling matrices is not only
algebraically more involved but also qualitatively more interesting
as reflected, for instance, in the RG equations.

In section \ref{sec:2model} we consider a model with spontaneously
broken $SU(N)$ symmetry which is, in fact, the matter sector of an
$SU(N)$ Georgi-Glashow model coupled to fermions.  This is quite
different from the model in the previous section, since mixing is
controlled by the continuous symmetry and its patterns of spontaneous
and explicit breaking, and it occurs mostly due to the quartic
self-interactions.  We restrict ourselves to \msb\ in that section and
to one particular type of spontaneous symmetry breaking (SSB).  We
discuss in some detail the simpler case in which there is no explicit
$SU(N)$ breaking, and consequently no coupling matrix, and comment
briefly on the case with the symmetry explicitly broken by mass terms,
in which an orthogonal coupling matrix enters the Lagrangian.  In
section \ref{sec:finale} we give our final remarks.  Complementary
material is gathered in several appendices.

\section{A model of scalar mixing with \ns\ scalars and \nf\ fermions} 
\label{sec:1model}

We consider a schematic model involving \ns\ real scalar fields
$\mat{\phi}=(\phi_\sss{1},\ldots,\phi_\sss{\ns})$ and $\nf$ fermions
$\mat{\psi}=(\psi_\sss{1},\ldots,\psi_\sss{\nf})$ interacting through
a Yukawa coupling.  The Lagrangian includes an additional field $\chi$
not coupled to fermions, chosen to be a scalar for simplicity,
$\Ll = \Ll_\sss{\mathrm{free}} +  \Ll_\sss{\mathrm{int}}$ with
\begin{equation}\label{eq:lag}
\begin{aligned}
%
  \Ll_\sss{\mathrm{free}} &= -\frac{1}{2} \chi \left(\Box +
    m^2_\sss{\chi}\right) \chi -\frac{1}{2} \mat{\phi} \pdot \left( \Box +
    \mat{M^2}\right) \mat{\phi} + \overline{\mat\psi}\, i\!\dirac\,
  \mat\psi, \\
  \Ll_\sss{\mathrm{int}} &= -\frac{\xi}{3!} \chi^3 - \frac{\lambda}{4!} \chi^4
  - V_4(\mat\phi) -\frac{1}{2} \mat\phi \pdot \mat{H} \mat\phi\, \chi -
  \sum_{k=1}^{\nf} \mat{a}_\sss{k} \pdot \mat\phi
  \,\overline{\psi}_\sss{k} \psi_\sss{k} ~.
\end{aligned}
\end{equation}
The \mat{\phi} mass matrix \mat{M^2} is real symmetric,
positive-definite and regular (i.e., none of its eigenvalues is
degenerate).  The mass--degenerate case is discussed separately below.
The \mat{\phi}--$\chi$ interaction term is bilinear in \mat{\phi},
with a real symmetric coupling matrix \mat{H}.  $V_4(\mat\phi)$
contains only terms quartic in \mat{\phi}, whose explicit form is
considered in detail below.  Yukawa interactions are specified by
$\nf$ real coupling vectors $\mat{a}_\sss{k} = (a_\sss{k
  1},\dots,a_\sss{k \ns})$, $k= 1,\dots,\nf$.  The field \mat{\phi}\ 
takes values in real \ns\ dimensional Euclidean space, which we refer
to as ``(scalar) flavor space.'' We do not use the summation
convention for the flavor indices of scalar and fermion fields.

$\Ll$ is invariant under an exact $U(1)^{\nf}$ symmetry of the fermion
sector, related to the conservation of the number of each fermion
species $\psi_\sss{k}$, which therefore do not mix among each other.
The fermion fields \mat{\psi} are chosen to be massless.  $\Ll$ is then
also invariant under the exact \zz\ symmetry $\mat{\phi} \rightarrow
-\mat{\phi}$, $\mat{\psi}\rightarrow \gamma_\sss{5} \mat{\psi}$, which
forbids vertices involving an odd number of \mat{\phi} fields,
couplings of $\chi$ to fermions, and $\chi$--$\mat{\phi}$ mixing.  An
interaction term of the form $\mat{\phi} \pdot \mat{K}\mat{\phi}
\chi^2$ with a coupling matrix \mat{K} is allowed by the \zz\ 
symmetry, but we do not include it in $\Ll$ for simplicity.

Let us call $\sy$ the subspace of flavor space spanned by
$\{\mat{a}_\sss{k}\}_{k=1}^{\nf}$ and $\dy$ its dimension, so $\dy
\leq \nf$ and $\dy \leq \ns$.  If $\dy < \ns$, the Yukawa interaction
is invariant under the discrete symmetry \zy\ that obtains by
restricting the transformations of the exact \zz\ symmetry of \Ll\ 
described above to the components of \mat{\phi} lying on \sy.  In
other words, under \zy\ only the components of \mat{\phi} coupled to
fermions change sign.  If $\dy \leq \ns - 2$, the Yukawa interaction
is obviously invariant under the group $SO(\ns-\dy)$ acting in the
orthogonal complement of \sy.  We assume that $V_4(\mat\phi)$ is
chosen so that it possesses the same symmetries as the Yukawa
interactions.  Thus, both \zy\ and $SO(\ns-\dy)$ are approximate
symmetries of \Ll, softly broken by $\chi$--$\mat{\phi}$ interactions
(unless \sy\ is contained in an eigenspace of \mat{H}) and by the
\mat{\phi} mass term (unless \sy\ is contained in an eigenspace of
\mat{M^2}).

Depending on the geometric configuration of $\{\mat{a}_\sss{k}\}$
there can be other, generally approximate symmetries of \Ll.  If
$\sy=\sya{1} \oplus \sya{2}$ with \sya{1} generated by, say, $\{
\mat{a}_\sss{j}, 1 \leq j \leq n_\sss{1}\}$ and orthogonal to \sya{2},
generated by $\{ \mat{a}_\sss{j}, n_\sss{1} < j \leq \nf\}$, we have
the approximate symmetry groups \zya{1}: $\mat{\hat{a}}_\sss{j} \pdot
\mat{\phi} \rightarrow -\mat{\hat{a}}_\sss{j} \pdot \mat{\phi}$,
$\mat{\psi}_\sss{j} \rightarrow \gamma_\sss{5} \mat{\psi}_\sss{j}$, if
$j \leq n_1$, with all other fields unmodified, and similarly \zya{2}:
$\mat{\hat{a}}_\sss{k} \pdot \mat{\phi} \rightarrow
-\mat{\hat{a}}_\sss{k} \pdot \mat{\phi}$, $\mat{\psi}_\sss{k}
\rightarrow \gamma_\sss{5} \mat{\psi}_\sss{k}$, if $n_1 < k \leq \nf$,
all other fields unaffected.  Here we denoted $\mat{\hat{a}}_\sss{i} =
\mat{a}_\sss{i}/||\mat{a}_\sss{i}||$.  These symmetries are softly
broken unless \sya{1} and \sya{2 } are included in eigenspaces of
\mat{H} and \mat{M^2}.  The converse is also true, if \zya{1} or
\zya{2} are exact symmetries of \Ll, then \sya{1}\ and \sya{2}\ must be
contained in eigenspaces of \mat{H} and \mat{M^2}.

Similar approximate discrete symmetries appear if the
$\{\mat{a}_\sss{k}\}$ lie on several different orthogonal subspaces,
and if within each subspace the coupling vectors are collinear.  One
extreme situation is when all of the coupling vectors
$\{\mat{a}_\sss{k}\}$ are pairwise orthogonal, with $\nf \leq \ns$. In
that case we have a $\zz^{\nf}$ symmetry, in general also softly
broken by \mat{M^2} and \mat{H}.  The other extreme is when all
coupling vectors are collinear, so that only one component of
\mat{\phi} is coupled to fermions.  In the special case in which all
of the $\mat{a}_\sss{k}$ are eigenvectors of \mat{H} there is an
interaction basis (if $V_4$ is appropriately chosen), and if they are
also eigenvectors of \mat{M^2} mixing becomes trivial.  Also, if out
of the $\nf$ coupling vectors $\mat{a}_\sss{k}$, $n$ (with $n \le
\nf$) are equal, and orthogonal to the remaining ones, the Lagrangian
is invariant under the corresponding $SU(n)$ symmetry of the fermion
sector.  The stability under renormalization of those special
geometric configurations of $\{\mat{a}_\sss{k}\}$ is of particular
interest. As expected, all of them are critical points of the
renormalization group (henceforth RG) flow.  We consider these issues
in detail below, in \msb\ scheme.

Despite its simplicity, in general the Lagrangian (\ref{eq:lag}) does
not have an interaction basis in which $\Ll_\sss{\mathrm{int}}$ is
diagonal in \mat{\phi}, except in the special cases mentioned above.
This feature is interesting to point out because some approaches to
mixing renormalization rely on the existence of a flavor basis
diagonalizing the interactions that is orthogonally, or unitarily,
related to the mass basis. In fact, we could further generalize our
model to have a Yukawa interaction of the form $\sum_{a=1}^{\ns}
\overline{\mat{\psi}} \mat{A}_\sss{a} \mat{\psi} \phi_\sss{a}$, where
$\{\mat{A}_\sss{a}\}$ are $\ns$ Hermitian $\nf \times \nf$ matrices.
$\Ll$ in (\ref{eq:lag}) corresponds to the case where all
$\mat{A}_\sss{a}$ are simultaneously diagonal.  In the general case
where not all those matrices commute there is no diagonal ``weak''
basis for the fermion fields either, even though they are massless.

\subsection{Quartic terms in the potential}
\label{sec:pot}

The simplest form for the quartic interactions $V_4(\mat\phi)$ is
isotropic in flavor space.  That form is not enough to make \Ll\ 
renormalizable, so we adopt the following parametrization for $V_4$,
\begin{equation}
  \label{eq:pot}
  V_4(\mat\phi) = \frac{g}{4!} (\mat{\phi} \pdot \mat{\phi})^2 +
  \frac{1}{4!} \sum_{i \leq j=1}^{\nf} g_{ij} (\mat{\hat{a}}_i \pdot
  \mat{\phi}) (\mat{\hat{a}}_j \pdot \mat{\phi}) (\mat{\phi} \pdot
  \mat{\phi}) +
  \frac{1}{4!} \sum_{i \leq j \leq k \leq l =1}^{\nf} g_{ijkl} (\mat{\hat{a}}_i \pdot
  \mat{\phi}) (\mat{\hat{a}}_j \pdot \mat{\phi}) (\mat{\hat{a}}_k \pdot
  \mat{\phi})  (\mat{\hat{a}}_l \pdot \mat{\phi}).
\end{equation}
We assume that $g$, $g_{ij}$ and $g_{ijkl}$ are such that
$V_4(\mat\phi) \geq 0$ for all \mat{\phi}.  If $\ns > \dy +1$,
$V_4(\mat\phi)$ is invariant under the $SO(\ns-\dy)$ transformations
leaving all components $\mat{a}_k \pdot \mat{\phi}$ invariant.  The
coupling constants $g_{ij}$ and $g_{ijkl}$ are chosen so that
$V_4(\mat\phi)$ is invariant under all of the discrete symmetries of
the Yukawa interactions.  Thus, in the example given above of two
orthogonal subspaces \sya{1} and \sya{2}, we must have $g_{ij} = 0 =
g_{ijkl}$ if an odd number of indices take values $\leq n_1$.  To give
another example, if the $\mat{a}_k$ are pairwise orthogonal then only
the couplings $g$, $g_{ii}$ and $g_{iijj}$ with $i \leq j$ may not 
vanish. 

\subsection{Quantum Lagrangian}
\label{sec:qlag}

Our perturbation expansion parameters are $\lambda$, $a_\sss{k} \equiv
||\mat{a}_\sss{k}||$, $g$, $g_{ij}$, $g_{ijkl}$ and the dimensionful
$\xi$ and \mat{H}.  We introduce $\chi$ wave-function and mass
renormalization constants $\delta Z_\chi$ and $\delta m_\chi^2$ in the
usual manner.  Let us assume for the moment that an interaction term
$\Ll_\sss{K} = - \mat{\phi} \pdot \mat{K} \mat{\phi} \chi^2$ is added
to $\Ll_\sss{\mathrm{int}}$, with \mat{K}\ positive definite.  It is
easily seen that \mat{K}\ must be multiplicatively renormalizable.  In
$\mathrm{MS}$ scheme the counterterms to \mat{K}\ must be of the form
\begin{equation}
  \label{eq:kterm}
\delta K_{ab} = \mu^{-\epsilon} \left(\sum_{cd} \delta
  Z_{abcd}^\sss{(K)} K_{cd} + \delta Z_{ab}^{\sss{(K)}\prime}\right), 
\end{equation}
with $\epsilon = 4-d$.  The second term in (\ref{eq:kterm}) collects
divergent $\epsilon$-pole contributions to $\delta \mat{K}$ that do
not vanish if $\mat{K} = 0$.  Since $\delta Z^{\sss{(K)}\prime}$
originates from diagrams with two \mat{\phi}- and two $\chi$- external
lines and no factor of \mat{K}, it must vanish when $\mat{H}=0$.  But
$\delta Z^{\sss{(K)}\prime}$ cannot depend on \mat{H} \cite{col1},
therefore it vanishes.  Thus, if we set $\mat{K}=0$ as done in
(\ref{eq:lag}), all Green's functions of $\phi_a \phi_b \chi^2$ are
finite once the other couplings and the propagators have been
renormalized (with \mat{K}=0).

There is no big loss of generality in setting $\mat{K}=0$.  If we had
a term $\Ll_\sss{K}$ in $\Ll$ we would set $\phi_\sss{\ns +1}\equiv\chi$
and redefine $V_4 \rightarrow V_4 - \Ll_\sss{K}$.  We would still have
the exact \zz\ symmetry forbidding additional cubic interactions and
$\chi$--$\phi$ mixing.  We would require, however, that $\Ll_\sss{K}$
have the same discrete symmetries as the Yukawa term.

Having set $\mat{K}=0$, a similar argument goes through for $\lambda$.
We can then just set $\lambda=0$ in $\Ll$ without spoiling its
renormalizability.  If we also set $\xi=0$, Green's functions of
$\chi^3$ receive contributions only from triangle-type diagrams with
three $\mat{\phi} \cdot \mat{H} \mat{\phi} \chi$ external vertices.
Such diagrams have negative superficial degree of divergence and are
therefore finite in the renormalized theory with $\xi =0$.  We arrive
in this way at a minimal $\chi$ sector, which still serves its purpose
of inducing bilinear mixing of \mat{\phi}.

The fermion fields do not mix, so their wave-function renormalization
matrix is diagonal, $\psi_\sss{k_\naught} = Z^{1/2}_\sss{\psi_k}
\psi_\sss{k} = (1+1/2 \delta Z_\sss{\psi_k}) \psi_\sss{k}$.  The
renormalized scalar field \mat{\phi}\ can be related to the bare one
as $\mat{\phi}_\naught = \mat{A} \mat{\phi}$, with \mat{A}\ a real
$\ns \times \ns$ matrix.  It will prove convenient to introduce the
real polar decomposition of \mat{A}\ explicitly \cite{bou},
\begin{equation}
  \label{eq:wfr}
  \mat{\phi}_\naught = \mat{Q} \mat{Z}^\frac{1}{2} \mat{\phi}, \quad
  \mat{Q} = e^{-\mat{\delta Q}}, \quad
  \mat{Z}^\frac{1}{2} = \mat{1} + \frac{1}{2} \mat{\delta Z}, 
\end{equation}
where \mat{Q}\ is orthogonal, \mat{Z}\ symmetric and positive
definite, $\mat{\delta Q} = -\mat{\delta Q}^t$ and $\mat{\delta Z} =
\mat{\delta Z}^t$.  In \ms\ scheme we set $\mat{\delta Q}=0$ by
definition and, owing to the softly broken \zy\ symmetry, \mat{\delta
  Z} must have the form,
\begin{equation}
  \label{eq:wfrs}
  \mat{\delta Z} = \sum_{k=1}^{\nf} \delta Z_\sss{\phi k}
  \mat{\hat{a}}_\sss{k} \otimes \mat{\hat{a}}_\sss{k} +
  \sum_{j<k=1}^{\nf} \delta Z_\sss{\phi jk}
  \left( \mat{\hat{a}}_\sss{j} \otimes \mat{\hat{a}}_\sss{k} +
    \mat{\hat{a}}_\sss{k} \otimes \mat{\hat{a}}_\sss{j} \right) +
  \mat{\delta Z^\prime},
  \quad\text{with}\quad
  \mat{\delta Z^\prime} \mat{\hat{a}}_\sss{i} = 0, ~
  i \leq \nf.
\end{equation}
Furthermore, if $\ns > \dy + 1$, due to the softly broken
$SO(\ns-\dy)$ symmetry, \mat{\delta Z^\prime}\ must be diagonal.  In
the superrenormalizable case $\mat{a}_\sss{k} = 0 = V_4$ we must also
have $\mat{\delta Z}=0$ in \ms, since there is no divergent field
renormalization, but even in that case $\mat{\delta Q}$ is divergent
in \os\ scheme due to infinite mass renormalization.

We write the relation between bare and renormalized \mat{\phi}\ mass
matrices as \cite{bou},
\begin{equation}
  \label{eq:mr}
  \mat{M^2}_\naught = \mat{Q_m}^t \left( \mat{M^2} + \mat{\delta M^2}
  \right) \mat{Q_m}, \quad
  \mat{Q_m} = e^{-\mat{\delta Q_m}}, \quad
  \left[ \mat{\delta M^2}, \mat{M^2} \right] = 0,
\end{equation}
with \mat{Q_m}\ orthogonal, $\mat{\delta Q_m}^t = -\mat{\delta Q_m}$,
and $(\mat{\delta M^2})^t = \mat{\delta M^2}$.  At one-loop level we
have $\mat{M^2}_\naught = \mat{M^2} + \mat{\delta M^2} + [\mat{\delta
  Q_m}, \mat{\delta M^2}]$, with \mat{\delta Q_m} determined up to
addition of an antisymmetric matrix commuting with \mat{M^2}.
\mat{Q_m} is then defined up to multiplication by an element of the
subgroup of $SO(\ns)$ that leaves \mat{M^2}\ invariant.  We write the
counterterms to \mat{H}\ similarly \cite{bou}, taking into account
that it must be multiplicatively renormalizable,
\begin{equation}
  \label{eq:hr}
  \mat{H}_\naught = \mu^{\epsilon/2} \mat{V}^t \mat{Z_H} \mat{H}
  \mat{V}, \quad
  \mat{V} = e^{-\mat{\delta V}}, \quad
  [\mat{Z_H},\mat{H}] = 0,
\end{equation}
where \mat{V}\ is orthogonal, $\mat{\delta V} = -\mat{\delta V}^t$,
$\mat{Z_H} = \mat{1} + \mat{\delta Z_H} = \mat{Z_H}^t$.  \mat{V}\ 
depends not only on \mat{H}\ and the coupling constants $g$, $g_{ij}$,
$g_{ijkl}$, but also on the unit vectors $\mat{\hat{a}}_\sss{k}$,
through $V_4$.  For this reason, we do not expect to be able to set
$\mat{\delta V} =0$, not even in MS scheme, unlike the case in
\cite{bou}.  At one loop we have $\mat{H}_\naught = [\mat{\delta V},
\mat{H}] + \mat{\delta Z_H} \mat{H}$, with \mat{\delta V} determined
up to addition of an antisymmetric matrix commuting with \mat{H}.

The counterterm parametrizations (\ref{eq:mr}), (\ref{eq:hr}) are a
restriction to real matrices of the one given in \cite{bou} for
complex normal matrices.  As such, it is applicable in general to real
normal matrices (i.e., matrices \mat{A} such that $[\mat{A},
\mat{A}^t] = 0$.)  For such matrices a multiplicative transformation
by congruence $\mat{A}_\naught = \mat{B}^t \mat{A} \mat{B}$ is not
possible if we want both \mat{A}\ and its transformed
$\mat{A}_\naught$ to be normal, except in the particular case when
both are symmetric.  On the other hand, if \mat{A} and
$\mat{A}_\naught$ are normal but not symmetric, it may not be
convenient to split them in their symmetric and antisymmetric parts
and to treat them separately, as is clearly the case for orthogonal
coupling matrices.

Yukawa couplings are multiplicatively renormalizable.  Since fermions
do not mix, if we set some $\mat{a}_n=0$ the corresponding fermion
field $\psi_n$ completely decouples.  The relation between bare and
renormalized coupling vectors is of the form $\mat{a}_{j_\naught} =
\mu^{\epsilon/2} \mat{A}_j \mat{a}_j$, with $\mat{A}_j$ a real
$\ns\times\ns$ matrix depending on $\{\mat{a}_k\}$ and, in principle,
all other couplings in $\Ll$.  Any such relation between two vectors
can be written as the product of a rotation taking $\mat{\hat{a}}_j$
into $\mat{\hat{a}}_{j_\naught}$, times a dilatation changing $a_j$ to
$a_{j_\naught}$,
\begin{equation}
  \label{eq:yuk}
  \mat{a}_{k_\naught} = \mu^{\epsilon/2} Z_{a_k} \mat{Q}_{a_k}
  \mat{a}_k, \quad
  Z_{a_k} > 0, \quad
  \mat{Q}_{a_k} = e^{-\mat{\delta Q}_{a_k}}, \quad
  \mat{\delta Q}_{a_k} = -\mat{\delta Q}_{a_k}^t.  
\end{equation}
This separate treatment of the renormalization of $a_k \equiv
||\mat{a}_k||$ and $\mat{\hat{a}}_k \equiv \mat{a}_k/a_k$ is necessary
for the renormalizaton of the potential and turns out to be convenient
for the derivation of RG equations.  At one loop, $\mat{a}_{k_\naught}
= \mu^{\epsilon/2} (\mat{a}_k + \mat{\delta a}_k)$ with $\mat{\delta
  a}_k = \delta Z_{a_k} \mat{a}_k - \mat{\delta Q}_{a_k} \mat{a}_k$.
The softly broken \zy\ symmetry of $\Ll$ prevents the subspace \sy\ 
from receiving infinite renormalization.  Thus, in \ms\ we expect the
counterterms $\mat{\delta a}_k$ to be linear combinations of
$\{\mat{a}_l\}$.  Geometric configurations giving rise to further
approximate symmetries, as described above, are also preserved in this
scheme.  In \os\ scheme the symmetry breaking due to
$\chi$--$\mat{\phi}$\ interactions and \mat{\phi}\ mass term is
apparent in the counterterms, with $\mat{\delta a}_k$ having finite
components orthogonal to \sy.

We consider, finally, the renormalization of quartic \mat{\phi}\ 
self-couplings.  The coupling constants $g$, etc., in $V_4$ receive
infinite additive renormalization at one loop from fermion box
diagrams.  We set,
\begin{equation}
  \label{eq:gs}
  g_\naught = \mu^\epsilon (g + \delta g), \quad
  g_{ij_\naught} = \mu^\epsilon (g_{ij} + \delta g_{ij}), \quad
  g_{ijkl_\naught} = \mu^\epsilon (g_{ijkl} + \delta g_{ijkl}).
\end{equation}
The complete set of counterterms to $V_4$ is obtained by expanding
\begin{multline}\label{eq:cpot}
  V_4 \left(\mat{\phi}_\naught, \{\mat{\hat{a}}_{k_\naught}\},
    g_\naught, \{g_{ij_\naught}\}, \{g_{ijkl_\naught}\}\right) = \\
  V_4 \left(\mat{Q} \mat{Z}^{1/2} \mat{\phi}, \{\mat{Q}_{a_k}
    \mat{\hat{a}}_{k}\}, \mu^\epsilon (g + \delta g),\{\mu^\epsilon
    (g_{ij} + \delta g_{ij})\}, \{\mu^\epsilon (g_{ijkl} + \delta
    g_{ijkl})\}\right)
\end{multline}
in powers of the coupling constants (see appendix \ref{sec:v4ct}). 

We interpret the classical Lagrangian $\Ll$ in (\ref{eq:lag}) as being
written in terms of bare fields, mass and coupling constants, and
substitute in it the above expressions for bare quantities.  The
resulting Lagrangian, expressed in terms of renormalized quantities,
does not depend on the orthogonal matrices \mat{Q}, \mat{Q_m}, \mat{V}
and $\{\mat{Q}_{a_k}\}$ independently, but only on \mat{Q_m}\mat{Q},
\mat{V}\mat{Q} and $\{\mat{Q}_{a_k}^t\mat{Q}\}$.  This ambiguity is
fixed by imposing appropriate renormalization conditions.  In $\ms$
and similar schemes we set $\mat{Q}=\mat{1}$, whereas in
\os\ scheme we choose to parametrize the theory in such a
way that $\mat{M^2}_\naught$ and \mat{M^2} are simultaneously diagonal
by setting $\mat{Q_m}=\mat{1}$.

The Lagrangian is then given in terms of renormalized fields and
parameters by (\ref{eq:lag}) and (\ref{eq:pot}), with $\lambda = 0 =
\xi$ for simplicity, and with the substitutions $\mat{H} \rightarrow
\mu^{\epsilon/2} \mat{H}$, $\mat{a}_k \rightarrow \mu^\epsilon
\mat{a}_k$, $g \rightarrow \mu^\epsilon g$, etc., plus the counterterm
Lagrangian $\Ll_\sss{\mathrm{ct}}$.  Retaining only those counterterms
that contribute at the one-loop level, $\Ll_\sss{\mathrm{ct}}$ has the
form,
\begin{multline}
  \label{eq:ctl}
  \Ll_\sss{\mathrm{ct}} = -\frac{1}{2} \delta Z_\chi \chi \Box \chi -
  \frac{1}{2} \delta m_\chi^2 \chi^2 - \frac{1}{2} \mat{\phi} \pdot
  \mat{\delta Z} \Box \mat{\phi} - \frac{1}{2} \mat{\phi} \pdot
  \mat{\Delta M^2} \mat{\phi} \\
  +\sum_{k=1}^{\nf} \delta Z_k \overline{\psi}_k \, i\!\dirac\, \psi_k -
  \frac{1}{2} \mu^{\epsilon/2} \mat{\phi} \pdot \mat{\Delta H}
  \mat{\phi} \chi - \mu^{\epsilon/2} \sum_{k=1}^{\nf} \mat{\Delta a}_k
  \pdot \mat{\phi} \overline{\psi}_k \psi_k + \delta V_4
\end{multline}
with, 
\begin{equation}
  \label{eq:counter}
  \begin{split}
    \mat{\Delta M^2} & = \mat{\delta M^2} + \left[ \mat{\delta Q} +
    \mat{\delta Q_m}, \mat{M^2} \right] + \frac{1}{2} \{\mat{\delta
    Z}, \mat{M^2}\}, \\
   \mat{\Delta H} & = \left[ \mat{\delta V} + \mat{\delta Q}, \mat{H}
    \right] + \frac{1}{2} \left\{ \mat{\delta Z}, \mat{H} \right\} +
    \mat{\delta Z_H} \mat{H} + \frac{1}{2} \delta Z_\chi \mat{H},\\
    \mat{\Delta a}_k & = \left( \delta Z_{a_k} + \delta Z_{\psi_k} -
    \mat{\delta Q}_{a_k} + \mat{\delta Q} + \frac{1}{2} \mat{\delta Z}
    \right) \mat{a}_k, 
  \end{split}
\end{equation}
and $\delta V_4$ as obtained from (\ref{eq:cpot}) in appendix
\ref{sec:v4ct}.

\subsection{On-shell scheme}
\label{sec:oss}

In \os\ scheme we set $\mat{\delta Q_m}=0$, and require \mat{M^2} to
be diagonal to all orders in perturbation theory.  We assume \mat{M^2}
to be regular and then, since $[\mat{\delta M^2},\mat{M^2}]=0$ by
definition, \mat{\delta M^2} and $\mat{M^2}_\naught$ must both be
diagonal.  We assume for simplicity that $\chi$ and $\mat{\phi}$
masses are such that $\chi$ is stable, and that decays $\phi_a
\rightarrow \phi_b+ \chi$ are not allowed.  The masses of stable
particles, fermions and $\chi$, are pole masses in this scheme.
\mat{\phi} states are unstable.  We adopt the \os\ renormalization
scheme for the \mat{\phi} propagator as well (see e.g.\ \cite{den}).

The $\chi$ field 2-point function $\Gamma_{2\chi}(p^2) = p^2-m_\chi^2
- \Pi_\chi(p^2)$ is obtained by imposing the renormalization
conditions $\Pi_\chi(m_\chi^2) = 0 = \Pi_\chi^\prime(m_\chi^2)$, where
the prime stands for $\partial/\partial p^2$.  Denoting
$\Omega_\chi(p^2)$ the unrenormalized $\chi$ self-energy at one loop,
\begin{equation}
  \label{eq:xse}
  \Omega_\chi(p^2) = - \frac{1}{16 \pi^2 \epsilon} \tr\left( \mat{H}^2
  \right) +
  \frac{1}{32\pi^2} \sum_{a,b=1}^{\ns} H_{ab} H_{ba} b_0(p^2,m_a^2,m_b^2),
\end{equation}
we get, 
\begin{gather}
  \label{eq:x2p}
  \delta Z_\chi = \Omega_\chi^\prime(m_\chi^2), \quad
  \delta m_\chi^2 = m_\chi^2 \Omega_\chi^\prime(m_\chi^2) -
  \Omega_\chi(m_\chi^2) \\
  \Gamma_{2\chi}(p^2) = \left( 1 + \Omega_\chi^\prime (m_\chi^2)
  \right) \left( p^2-m_\chi^2 \right) - \left( \Omega_\chi(p^2) -
  \Omega_\chi(m_\chi^2) \right),
\end{gather}
with $\Omega_\chi^\prime(m_\chi^2) < 0$ if $\chi$ is stable, and
$\Gamma_{2\chi}(p^2)$ independent of the dimensional regularization
scale $\mu$.  In (\ref{eq:xse}), $b_0$ refers to the finite part of
the 2-point loop integral, as defined in appendix \ref{sec:apploop}.
As expected, $\delta Z_\chi$ is finite.

The \mat{\phi} 2-point function \[ \mat{\Gamma_2}(p^2) = p^2 \mat{1} -
\mat{M^2} - \mat{\Pi}(p^2) \] in this scheme is obtained with the
renormalization conditions,
\begin{equation}
  \label{eq:rencon}
  \Re \Pi_{aa}(m_a^2) = 0 = \Re \Pi_{aa}^\prime(m_a^2), 
  \qquad
  \Re \Pi_{ab}(m_a^2) = 0 = \Re \Pi_{ab}(m_b^2) ~ \mathrm{if} ~ a \neq
  b. 
\end{equation}
Here, the renormalized self-energy function $\mat{\Pi}(p^2) =
\mat{\Omega}(p^2) - p^2 \mat{\delta Z} + \mat{\Delta M^2}$ is given in
terms of the unrenormalized one,
\begin{equation}
  \label{eq:omega}
  \begin{split}
    \mat{\Omega}(p^2) & = \mat{\Omega}^{(1)}(p^2) +
    \mat{\Omega}^{(2)}(p^2) + \mat{\Omega}^{(3)}, \\
    \Omega^{(1)}_{ab}(p^2) & = -\frac{1}{8\pi^2\epsilon}
    (\mat{H}^2)_{ab} + \frac{1}{16\pi^2} \sum_{n=1}^{\ns} H_{an}
    H_{nb} b_0(p^2,m_\chi^2,m_n^2), \\
    \mat{\Omega}^{(2)}(p^2) & = -\frac{1}{4\pi^2\epsilon} p^2
    \sum_{k=1}^{\nf} \mat{a}_k \otimes \mat{a}_k +\frac{1}{8\pi^2} p^2 
    b_0(p^2,0,0) \sum_{k=1}^{\nf} \mat{a}_k \otimes \mat{a}_k  \\
  \end{split}
\end{equation}
where
\[
   b_0(p^2,0,0) = \log\left( -\frac{p^2}{\mub^2} -
  i\varepsilon \right) -2 = \log\left( \frac{|p^2|}{\mub^2}
  \right) - i \pi \Theta(p^2) - 2,
\]
the second equality being valid for real values of $p^2$.  The term
$\mat{\Omega}^{(3)}$ in (\ref{eq:omega}) gathers the contributions to
$\mat{\Omega}(p^2)$ from diagrams with one insertion of $V_4$.  These
are independent of $p^2$, real, and symmetric with respect to flavor
indices (we give their explicit expressions for the case $\ns=2$ in
appendix \ref{sec:appv4}).  $\mat{\Omega}(p^2)$ is then symmetric in
flavor space.

Applying the renormalization conditions (\ref{eq:rencon}) we get, for
the diagonal self-energies,
\begin{subequations}\label{eq:selfos}
\begin{gather}
  \Pi_{aa}(p^2)  = \Omega_{aa}(p^2) - \Re \Omega_{aa}(m_a^2) -
  (p^2-m_a^2) \Re \Omega^\prime_{aa}(m_a^2), \\
  \delta Z_{aa}  = \Re \Omega^\prime_{aa}(m_a^2),
  \quad
  \Delta M^2_{aa} = m_a^2 \Re \Omega^\prime_{aa} (m_a^2) - \Re
  \Omega_{aa} (m_a^2), \\
\intertext{and for the off-diagonal ones, $a\neq b$, }
  \Pi_{ab}(p^2)  = \Omega_{ab}(p^2) - \frac{p^2-m_b^2}{m_a^2-m_b^2}
  \Re \Omega_{ab}(m_a^2) + \frac{p^2-m_a^2}{m_a^2-m_b^2}
  \Re \Omega_{ab}(m_b^2), \\
  \delta Z_{ab}  = \frac{\Re \Omega_{ab}(m_a^2) - \Re
    \Omega_{ab}(m_b^2)}{m_a^2 - m_b^2},
  \quad
  \Delta M^2_{ab}  = \frac{m_b^2 \Re \Omega_{ab}(m_a^2) - m_a^2 \Re
    \Omega_{ab}(m_b^2)}{m_a^2 - m_b^2}.
\end{gather}
\end{subequations}
One-loop diagrams with a $V_4$ vertex do not contribute to
wave-function renormalization, so that \mat{\delta Z} is independent
of $\mat{\Omega}^{(3)}$.  Equation (\ref{eq:selfos}) then gives an
explicit expression for \mat{\delta Z}, which is of course symmetric,
and finite if all $\mat{a}_k = 0$.  The expression for $\mat{\Delta
  M^2 }$ depends on the detailed form of $\mat{\Omega}^{(3)}$.
From
(\ref{eq:counter}) with $\mat{\delta Q_m} =0$ we get
\[
\mat{\delta M^2} + \left[\mat{\delta Q}, \mat{M^2}\right] =
\mat{\Delta M^2} - \frac{1}{2} \left\{\mat{\delta Z}, \mat{M^2}\right\}.
\] 
Substituting the counterterms on the r.h.s.\ by their values from
(\ref{eq:selfos}), and taking into account that  \mat{\delta M^2} is
diagonal in this scheme and \mat{\delta Q} antisymmetric, we obtain,
\begin{equation}    \label{selfos2}
  \begin{split}
    \delta M^2_{aa} &= -\Re \Omega_{aa}(m^2_a), \qquad
    \delta M^2_{ab} = 0 \quad \mathrm{if} \quad a \neq b, \\
    \delta Q_{ab} &= \frac{1}{2} \frac{\Re \Omega_{ab}(m_a^2) + \Re
    \Omega_{ab}(m^2_b)}{m^2_a-m^2_b} \quad \mathrm{if} \quad a \neq b.
  \end{split}
\end{equation}
Notice that \mat{\delta Q} is divergent even if all $\mat{a}_k = 0$
(i.e., $\mat{\Omega}^{(2)}=0$) and $V_4 =0$ (i.e., $\mat{\Omega}^{(3)}
= 0$), due to divergent terms in $\mat{\Omega}^{(1)}$, although in
that case the theory is superrenormalizable.

The expression (\ref{selfos2}) for \mat{\delta Q}\ is valid only for
regular mass matrices.  If some of the squared masses $m_a^2$ are
degenerate the derivation given above for \mat{\delta Q}\ does not
hold, since in that case the renormalization conditions
(\ref{eq:rencon}) for off-diagonal 2-point functions are not
independent.  The treatment of the renormalization of 2-point
functions \mat{\Gamma_2}\ in the mass-degenerate case is completely
analogous to that given in \cite{bou} in the case of fermion
fields.

The fermion two-point functions have the form $\Gamma_{2\psi_k} (p) =
\pirac - \pirac \Pi_{\psi_k}(p^2)$, with 
\begin{equation}
  \label{eq:fse}
  \Pi_{\psi_k}(p)^2 = -\delta Z_{\psi_k} - \frac{1}{16 \pi^2 \epsilon}
  a_k^2 + \frac{1}{32 \pi^2 p^2} \sum_{c=1}^{\ns} {(\mat{a}_k)_c}^2 (m_c^2
  a_0(m_c^2) + (p^2 - m_c^2) b_0(p^2,m_c^2,0)).
\end{equation}
The renormalization condition $\Gamma_{2\psi_k} (p) u_k(p) = 0$ if
$p^2 = 0$ is trivially satisfied, whereas the condition $1/\!\pirac\,
\Gamma_{2\psi_k} (p) u_k(p) = u_k(p)$ if $p^2 = 0$, or $\Pi_{\psi_k}
(p^2=0) = 0$, leads to,
\begin{equation}
  \label{eq:fse2}
  \begin{split}
    \delta Z_{\psi_k} & = \frac{-1}{16 \pi^2 \epsilon} a_k^2 +
    \frac{1}{64 \pi^2} a_k^2 + \frac{1}{32 \pi^2} \sum_{c=1}^{\ns}
    {(\mat{a}_k)_c}^2 a_0(m_c^2),\\
    \Pi_{\psi_k} (p^2) & = -\frac{1}{64 \pi^2} a_k^2 + \frac{1}{32
      \pi^2 p^2} \sum_{c=1}^{\ns} {(\mat{a}_k)_c}^2 (p^2-m_c^2)
    (b_0(p^2,m_c^2,0) - a_0(m_c^2)),
  \end{split}
\end{equation}
where we used $b_0(0,m_c^2,0) = a_0(m_c^2)$, and
$b^\prime_0(0,m_c^2,0) = -1/(2 m_c^2)$, as is easy to check with the
expressions given in appendix \ref{sec:apploop}.

We consider next the \mat{\phi}--$\chi$ vertex function, which has the
form,
\begin{equation}
  \label{eq:scvr}
    \Gamma_{ab} (p_1,p_2)  = -\mu^{\epsilon/2} H_{ab}
    -\mu^{\epsilon/2} \Delta H_{ab}  + \Gamma_{ab}^{(1)} (p_1,p_2) +
    \Gamma_{ab}^{(2)} (p_1,p_2).
\end{equation}
Here, $\Gamma_{ab}^{(1)} (p_1,p_2)$ is the contribution from the
scalar triangle diagram,
\begin{equation}
  \label{eq:scvr1}
  \Gamma_{ab}^{(1)}(p_1,p_2)  = -\mu^{\epsilon/2} \sum_{c,d=1}^{\ns} H_{ac}
  H_{cd} H_{db} C_0(p_1+p_2, p_2, m_\chi^2, m_d^2, m_c^2),  
\end{equation}
where $C_0$ is the finite scalar triangle integral, as given in
appendix \ref{sec:apploop}.  There are no other triangle diagrams at
one loop, since we set $\xi = 0$.  $\Gamma_{ab}^{(2)} (p_1,p_2)$
collects the contributions from diagrams with one $V_4$ vertex, which
are divergent and depend only on $p_1 + p_2$.  Its explicit form in
the case $\ns=2$ is given in appendix \ref{sec:appv4}.  It is apparent
from (\ref{eq:scvr1}) that $\Gamma_{ab}^{(1)}(p_1,p_2) =
\Gamma_{ba}^{(1)}(p_2,p_1)$, and from appendix \ref{sec:appv4} that
$\Gamma_{ab}^{(2)}(p_1,p_2) = \Gamma_{ba}^{(2)}(p_2,p_1)$, therefore,
$\Gamma_{ab}(p_1,p_2) = \Gamma_{ba}(p_2,p_1)$.

If the external momenta in $\Gamma_{ab}(p_1,p_2)$ are on mass shell,
$p_1^2 = m_a^2$,   $p_2^2 = m_b^2$, and $(p_1+p_2)^2 = m_\chi^2$.
Defining 
\begin{equation}
  \label{eq:scvros}
  \Gamma_{ab}^{(\os)} = \left. \Gamma_{ab} (p_1,p_2) \right|_\mathrm{on-shell},
\end{equation}
we see that $\Gamma_{ab}^{(\os)}$ does not depend on momenta, since it
can only depend on $p_1^2$, $p_2^2$, $p_1 \pdot p_2$ which are fixed
by the on-shell conditions, and furthermore $\Gamma_{ab}^{(\os)} =
\Gamma_{ba}^{(\os)}$.  For concreteness, we impose on
$\Gamma_{ab}(p_1,p_2)$ the renormalization condition
$\Gamma_{ab}^{(\os)} = H_{ab}$, whence,
\begin{equation}
  \label{eq:osvr1}
  \Delta H_{ab}  =  \left[   \Gamma_{ab}^{(1)} (p_1,p_2)  +
  \Gamma_{ab}^{(2)} (p_1,p_2) \right]_\mathrm{on-shell}.
\end{equation}
The one-loop counterterm \mat{\Delta H} is completely fixed by this
equation.  We can, however, determine the counterterms to the coupling
matrix as defined in (\ref{eq:counter}) through the equation,
\begin{equation}
  \label{eq:heq}
  \mat{H} \mat{\delta Z_H} + \left[ \mat{\delta V}, \mat{H} \right] =
  \mat{\Delta H} - \left[ \mat{\delta Q}, \mat{H} \right] - \frac{1}{2}
  \left\{ \mat{\delta Z}, \mat{H} \right\} - \frac{1}{2} \delta Z_\chi
  \mat{H}
\end{equation}
where all the quantities on the r.h.s.\ are already known.  This
equation can always be solved for  \mat{\delta Z_H} and \mat{\delta
  V}, as shown in appendix C of \cite{bou}, by projecting over an
appropriate basis of matrices.  In appendix \ref{sec:appmat} we
consider the special case in which \mat{H} is regular.

We consider, finally, the renormalization of the Yukawa couplings, as
given by the 1-PI three-point Green's function, $\Gamma_{ka}
(p_1,p_2)$.  Here, $p_{1,2}$ are the momenta of the fermions, incoming
by convention, $k=1,\dots,\nf$ is the fermion flavor, and
$a=1,\dots,\ns$ the scalar flavor.  Only one fermion flavor index is
needed since, due to the $U(1)^{\nf}$ symmetry, Green functions of
$\phi_a \psi_k \overline{\psi}_j$ vanish unless $j=k$.  $\Gamma_{ka}
(p_1,p_2)$ is a bispinor which at tree level is proportional to the
identity matrix, $\Gamma_{ka} (p_1,p_2) = (\mat{a}_k)_a$.  Expanding
the one-loop $\Gamma_{ka} (p_1,p_2)$ in the usual Dirac matrix basis,
it is clear that only the scalar form factor can receive divergent
contributions, the other form factors being finite.  We focus then
on $\Gamma_{ka}^{(S)} (p_1,p_2) = 1/4 \mathrm{Tr}_D \Gamma_{ka}
(p_1,p_2)$, assuming for concreteness that the physical values of the
coupling vectors $\mat{a}_k$ are fixed by these form factors.

At one loop $\Gamma_{ka}^{(S)}$ is given by,
\begin{equation}
  \label{eq:yff}
  \Gamma_{ka}^{(S)}(p_1,p_2) = \mu^{\epsilon/2} (\mat{a}_k)_a +
  \mu^{\epsilon/2} (\mat{\Delta a}_k)_a + \mu^{\epsilon/2}
  (\mat{a}_k)_a \sum_{c=1}^{\ns} {(\mat{a}_k)_c}^2 f^{(S)} (p_1,p_2,m_c^2),
\end{equation}
where $f^{(S)}$ is the scalar form factor for the triangle diagram,
\begin{equation}
  \label{eq:tff}
  f^{(S)} (p_1,p_2,m_c^2) = \frac{-1}{8\pi^2 \epsilon} + \frac{1}{32
  \pi^2} \left( b_0(p_1^2,m_c^2,0) + b_0(p_2^2,m_c^2,0) \right) -
  \frac{(p_1+p_2)^2}{2} C_0(p_1+p_2,p_2,m_c^2,0,0).
\end{equation}
On the mass-shell $p_1^2=0=p_2^2$, $2 p_1 \pdot p_2 = m_a^2$.  Defining
\begin{equation}
  \label{eq:osyff}
  \Gamma_{ka}^{(\os)} =   \left. \Gamma_{ka}^{(S)}(p_1,p_2)
  \right|_\mathrm{on-shell},
  \qquad
  f^{(S)(\os)} (m_a^2,m_c^2) = \left. f^{(S)} (p_1,p_2,m_c^2)
  \right|_\mathrm{on-shell},
\end{equation}
and setting $\Re\Gamma_{ka}^{(\os)} = \mu^{\epsilon/2} (\mat{a}_k)_a$,
from (\ref{eq:yff}) we have,
\begin{equation}
  \label{eq:ffg}
  (\mat{\Delta a}_k)_a = -(\mat{a}_k)_a \sum_{c=1}^{\ns} {(\mat{a}_k)_c}^2
  \Re f^{(S)(\os)} (m_a^2,m_c^2). 
\end{equation}
These counterterms are enough to renormalize the scalar form factors
at one-loop level.  The renormalization constants $\delta Z_{a_k}$ and
$\mat{\delta Q}_{a_k}$ defined in (\ref{eq:yuk}) and
(\ref{eq:counter}) can be extracted from the value of $\mat{\Delta
  a}_k$. For that purpose, it is convenient to rewrite $f^{(S)(\os)}$
as,
\begin{subequations}
  \begin{gather}
    \Re f^{(S)(\os)} (m_a^2,m_c^2) = -\frac{1}{8\pi^2 \epsilon} +
    f^{(1)} (m_c^2) + f^{(2)} (m_a^2,m_c^2),  \label{eq:ffg1}\\
    f^{(1)} (m_c^2)  = \frac{1}{16\pi^2} a_0(m_c^2),
\qquad
    f^{(2)} (m_a^2,m_c^2) = \frac{m_a^2}{2} \left. \Re C_0
    (p_1+p_2,p_2,m_c^2,0,0) \right|_\mathrm{on-shell}.
  \end{gather}
\end{subequations}
Notice that $f^{(1)}$ depends only on the internal mass $m_c^2$,
whereas $f^{(2)} (m_a^2,m_c^2)$ depends on the external mass $m_a^2$
as well.

Starting from the definition of $\mat{\Delta a}_k$ in
(\ref{eq:counter}), we have,
\begin{equation}
  \label{eq:ycnt }
  \left( \delta Z_{a_k} - \mat{\delta Q}_{a_k} \right) \mat{a}_k =
  \mat{\Delta a}_k - \delta Z_{\psi_k} \mat{a}_k - \mat{\delta Q}
  \mat{a}_k - \frac{1}{2} \mat{\delta Z} \mat{a}_k,
\end{equation}
where all the quantities on the r.h.s.\ have already been computed.
The terms in $\delta Z_{\psi_k}$ and $\mat{\delta Q}$ on the r.h.s.\ 
already have the form required to determine the l.h.s.. In order to
cast the other ones as a dilatation times a rotation, we proceed as in
appendix \ref{sec:appdec}.  Thus, for the wave-function
renormalization matrix we write,
\begin{equation}
  \label{eq:wfcy}
  \mat{\delta Z} \mat{a}_k = \left( \mat{\hat{a}}_k \pdot \mat{\delta Z}
  \mat{\hat{a}}_k \right) \mat{a}_k + \left( \left( \mat{\delta Z}
  \mat{\hat{a}}_k \right) \wedge \mat{\hat{a}}_k  \right) \mat{a}_k. 
\end{equation}
with the usual definition of the wedge product as $\mat{u} \wedge
\mat{v} \equiv \mat{u} \otimes \mat{v} - \mat{v} \otimes
\mat{u}$.  For the vertex counterterm we have $\left( \mat{\Delta a}_k
\right)_c  = \delta_1 a_k\, (\mat{a}_k)_c + (\mat{\delta}_2
\mat{a}_k)_c$ with, 
\begin{equation}
  \label{eq:vcty}
    \delta_1 a_k  = \frac{a_k^2}{8\pi^2 \epsilon} - \sum_{c=1}^{\ns}
    {(\mat{a}_k)_c}^2 f^{(1)} (m_c^2),\qquad
    (\mat{\delta}_2 \mat{a}_k)_c  = -(\mat{a}_k)_c \sum_{d=1}^{\ns}
    {(\mat{a}_k)_d}^2 f^{(2)} (m_c^2,m_d^2).
\end{equation}
Defining the diagonal matrix $(\mat{\delta B})_{ab} = -\delta_{ab}
\sum_{d=1}^{\ns} {(\mat{a}_k)_d}^2 f^{(2)} (m_a^2,m_d^2)$ we can write
$\mat{\delta}_2 \mat{a}_k$ as $\mat{\delta}_2 \mat{a}_k = \left(
  \mat{\hat{a}}_k \pdot \mat{\delta B} \mat{\hat{a}}_k \right)
\mat{a}_k + \left( \left( \mat{\delta B} \mat{\hat{a}}_k \right)
  \wedge \mat{\hat{a}}_k  \right) \mat{a}_k$, whence,
\begin{equation}
  \label{eq:whnc}
  \mat{\Delta a}_k = \left( \delta_1 a_k +  \mat{\hat{a}}_k \pdot \mat{\delta
  B} \mat{\hat{a}}_k \right) \mat{a}_k + \left( \left( \mat{\delta B}
  \mat{\hat{a}}_k \right) \wedge \mat{\hat{a}}_k  \right) \mat{a}_k.
\end{equation}
Notice that only $\delta_1 a_k$ is divergent. Gathering all
contributions together we get,
\begin{equation}
  \label{eq:gthr}
    \delta Z_{a_k} = \delta_1 a_k + \left( \mat{\hat{a}}_k \pdot
    \mat{\delta B} \mat{\hat{a}}_k \right) - \delta Z_{\psi_k} -
  \frac{1}{2} \mat{\hat{a}}_k \pdot \mat{\delta Z} \mat{\hat{a}}_k,
  \quad
  \mat{\delta Q}_{a_k} = \mat{\delta Q} - \left( \mat{\delta B}
    \mat{\hat{a}}_k \right) \wedge \mat{\hat{a}}_k  + \frac{1}{2}
    \left( \mat{\delta Z} \mat{\hat{a}}_k \right) \wedge
    \mat{\hat{a}}_k.   
\end{equation}
These constants determine the renormalization of $a_k$ and
$\mat{\hat{a}}_k$ separately, the latter being necessary for the
renormalization of $V_4$.

\subsection{\msb scheme}
\label{sec:mss}

In \msb scheme we set $\mat{\delta Q} = 0$ by definition.  We choose
our flavor basis so that at tree level the renormalized mass matrix
\mat{M^2} is diagonal.  This choice makes the tree-level propagators
diagonal which, although by no means mandatory, conveniently
simplifies the treatment.  We write $\mat{M^2} = \mat{M^{\prime 2}} +
\mat{\widehat{M}^2}$, with $\mat{M^{\prime 2}} = \diag (m_1^2, \dots,
m_{\ns}^2)$ containing the renormalized masses, and
\mat{\widehat{M}^2} the off-diagonal elements, which are of second
order in the coupling constants, and are treated as interaction terms.
As pointed out in \cite{bou}, since the mass matrix is symmetric,
we can always write $\mat{M^2}=\exp (\mat{E}) \mat{M^{\prime 2}} \exp
(-\mat{E})$, with $\mat{M^{\prime 2}}$ the diagonal matrix of
eigenvalues and $\exp (\mat{E})$ the orthogonal matrix of normalized
eigenvectors of $\mat{M^2}$.  \mat{E} is antisymmetric and, due to our
choice of tree-level flavor basis, it is of second order in the
coupling constants.  At one-loop level, $\mat{M^2} = \mat{M^{\prime 2}}
+ [\mat{E}, \mat{M^{\prime 2}}] + \cdots$, with $\mat{\widehat{M}^2} =
[\mat{E}, \mat{M^{\prime 2}}]$ having vanishing diagonal entries.

We rewrite the \mat{\phi} field Lagrangian as,
\begin{equation}
  \label{eq:philag}
  \Ll_\phi = -\frac{1}{2} \mat{\phi} \pdot (\Box + \mat{M^{\prime 2}})
  \mat{\phi} - \frac{1}{2} \mat{\phi} \pdot \mat{\widehat{M}^2} \mat{\phi}
  - \frac{1}{2} \mat{\phi} \pdot (\mat{\delta Z}\Box + \mat{\Delta M^2})
  \mat{\phi}
\end{equation}
with,
\begin{equation}
  \label{eq:philag1}
  \mat{\Delta M^2} = \mat{\delta M^2} + [\mat{\delta Q_m}, \mat{M^2}]
  + \frac{1}{2} \{\mat{\delta Z}, \mat{M^2}\}.
\end{equation}
Counterterms to the \mat{\phi} two-point function are obtained from
(\ref{eq:omega}) as,
\begin{equation}
  \label{eq:msctr}
  \mat{\delta Z} = -\frac{1}{4 \pi^2 \epsilon} \sum_{k=1}^{\nf}
  \mat{a}_k \otimes \mat{a}_k, \qquad
  \mat{\Delta M^2} = \frac{1}{8\pi^2\epsilon} \mat{H}^2 +
  \left(\mat{\Omega}^{(3)}\right)_\mathrm{div.}. 
\end{equation}
\mat{\delta Z} is invariant under $\mat{a}_k \rightarrow -\mat{a}_k$,
$k=1,\dots,\nf$, as expected and, in the notation of (\ref{eq:wfrs}),
$\delta Z_{\phi k} = - a_k^2/(4\pi^2\epsilon)$, $\delta Z_{\phi jk} =
0 = \mat{\delta Z^\prime}$.  With these values (\ref{eq:msctr}) for
\mat{\delta Z} and \mat{\Delta M^2} we can find the mass
renormalization constants \mat{\delta M^2} and \mat{\delta Q_m} from
(\ref{eq:counter}),
\begin{equation}
  \label{eq:mscounter}
  \delta M_{ab}^2 + \left[ \mat{\delta Q_m}, \mat{M^2}\right]_{ab} =
  \frac{1}{8\pi^2\epsilon} \left(\mat{H}^2\right)_{ab} +
  \left(\Omega^{(3)}_{ab}\right)_{\mathrm{div.}} +
  \frac{m_a^2+m_b^2}{8\pi^2\epsilon} \sum_{k=1}^{\ns} (\mat{a}_k)_a
  (\mat{a}_k)_b. 
\end{equation}
Unlike the case in \os\ scheme, in \msb \mat{\delta Q_m} is not needed
in order to renormalize the coupling constants.  We use below the
complete mass matrix renormalization as given by (\ref{eq:mscounter})
to derive the one-loop RG equation for \mat{M^2}.  Regarding the
choice of \mat{\widehat{M}^2} and its inherent ambiguities,
considerations completely analogous to those given in Section 2.4 of
\cite{bou} are valid, there is no need to repeat them here.

The \mat{\phi} self-energy in this scheme is then given by the
$\epsilon$-independent terms in (\ref{eq:omega}), augmented by the
off-diagonal mass matrix, $\mat{\Pi}(p^2) =
(\mat{\Omega}(p^2))_\mathrm{finite} + \mat{\widehat{M}^2}$.  The
relation among renormalized masses and pole parameters is obtained
from $\Gamma_{2aa}(s_{p_a}) = 0$, with $s_{p_a}=m_{p_a}^2 - i m_{p_a}
\Gamma_{p_a}$.  At one-loop level,
\begin{equation}
  \label{eq:mspole}
  m_a^2 = m_{p_a}^2 - \Re \Pi_{aa}(m_{p_a}^2), \qquad
  \Gamma_{p_a} = \frac{m_{p_a}}{8\pi} \sum_{k=1}^{\nf} {(\mat{a}_k)_a}^2.
\end{equation}
The one-loop self-energies and wave-function and mass counterterms for
$\chi$ and fermion fields are given by,
\begin{gather}
  \delta Z_{\psi_k} = -\frac{1}{16\pi^2 \epsilon} a_k^2,
  \qquad
  \Pi_{\psi_k} (p^2) = \frac{1}{32 \pi^2 p^2} \sum_{c=1}^{\ns}
  {(\mat{a}_k)_c}^2 \left( m_c^2 a_0(m_c^2) + (p^2-m_c^2)
    b_0(p^2,m_c^2,0) \right), \label{eq:mszpsi}\\
  \delta Z_{\chi} = 0,
  \qquad
  \delta m_\chi^2 = \frac{1}{16\pi^2 \epsilon} \Tr{\mat{H}^2},
  \qquad
  \Pi_\chi (p^2) = \frac{1}{32 \pi^2} \sum_{a,b=1}^{\ns} H_{ab} H_{ba}
  b_0(p^2,m_a^2,m_b^2), \label{eq:mszchi}
\end{gather}
from whence $m_\chi^2 = m_{\chi p}^2 - \Pi_\chi (m_{\chi p}^2)$.  This
expression satisfies the RG equation for $m_\chi^2$ obtained from its
counterterm, $\overline\mu d m_\chi^2 / d\overline\mu = 1/(16 \pi^2)
\Tr{\mat{H}^2}$, as can be checked from (\ref{eq:mszchi}).

The unrenormalized $\mat{\phi}$--$\chi$ vertex function has the form
given in (\ref{eq:scvr}).  The only divergent contributions to
$\Gamma_{ab}(p_1,p_2)$ at one loop come from diagrams with one $V_4$
vertex.  The detailed form of $\mat{\Delta H}$ in the case $\ns=2$ is
given in appendix \ref{sec:appv4}.  The renormalization constants
\mat{\delta Z_H} and \mat{\delta V} defined in (\ref{eq:hr}) satisfy
(\ref{eq:heq}) with $\mat{\delta Q}=0$.  If we set all $\mat{a}_k = 0$
and $V_4$ flavor-isotropic we can set $\mat{\delta V} = 0$, since in
that case \mat{\delta V} can only depend on \mat{H} and $g$ and
therefore commutes with \mat{H}.  In the case $\mat{a}_k \neq 0$ we
consider in this paper, $\mat{\delta V} \neq 0$ must be determined
from (\ref{eq:heq}).

We consider, finally, the renormalization of Yukawa couplings.  Since
$\mat{\delta Q} = 0$ and $\mat{\delta Z} = \sum_{j=1}^{\nf} \delta
Z^{(\phi)}_j \mat{\hat{a}}_j \otimes \mat{\hat{a}}_j$, we can write
the counterterm to the Yukawa interaction as,
\begin{equation}
  \label{eq:msyuk}
  \mat{\Delta a}_k = \left( \delta Z_{a_k} + \delta Z_{\psi_k} +
  \frac{1}{2} \delta Z^{(\phi)}_k  \right) \mat{a}_k - \mat{\delta
  Q}_{a_k} \mat{a}_k + \frac{1}{2} \sum_{j\neq k=1}^{\nf}
  \delta Z_j^{(\phi)} \left( \mat{\hat{a}}_j \pdot \mat{a}_k \right) 
  \mat{\hat{a}}_j. 
\end{equation}
In order for the interaction to retain its form under renormalization,
we must have $\mat{\delta Q}_{a_k} \mat{\hat{a}}_k \in \sy$.  This is
guaranteed in \msb by the softly broken \zy\ symmetry.  From the
expression for the unrenormalized Yukawa vertex scalar form factor
given above, we obtain, $\mat{\Delta a}_k = 1/(8\pi^2 \epsilon) a_k^2
\mat{a}_k$.  Substituting this value and the values from $\delta
Z_{\psi_k}$ and $\delta Z_j^{(\phi)}$ in (\ref{eq:msyuk}),
\begin{equation}
  \label{eq:msyct}
  \begin{split}
  \left( \delta Z_{a_k} - \mat{\delta Q}_{a_k}  \right) \mat{a}_k & = 
  \frac{5}{16 \pi^2 \epsilon} a_k^2 \mat{a}_k + \frac{a_k}{8 \pi^2
    \epsilon} \sum_{j\neq k=1}^{\nf} a_j^2 x_{jk} \mat{\hat{a}}_j,
  \\
   & = \frac{3}{16\pi^2 \epsilon} a_k^2 \mat{a}_k + \frac{1}{8\pi^2
  \epsilon} \sum_{j=1}^{\nf} a_j^2 x_{jk}^2 \mat{a}_k +
\frac{1}{8\pi^2 \epsilon} \left( \sum_{j=1}^{\nf} a_j^2 x_{jk}
  \mat{\hat{a}}_j \wedge \mat{\hat{a}}_k \right) \mat{a}_k,
  \end{split}
\end{equation}
where $x_{jk} \equiv \mat{\hat{a}}_j \pdot \mat{\hat{a}}_k$ and the
second line is obtained by substituting $\mat{\hat{a}}_j$ in the first
one by $ \mat{\hat{a}}_j = x_{kj} \mat{\hat{a}}_k + \left(
  \mat{\hat{a}}_j \wedge \mat{\hat{a}}_k \right) \mat{\hat{a}}_k.$
Clearly, $\mat{\delta Q}_{a_k}$ is determined only up to an
antisymmetric matrix nullifying $\mat{a}_k$.  From (\ref{eq:msyct}) we
see that we can set,
\begin{equation} \label{eq:msyct2}
  \delta Z_{a_k} = \frac{3}{16\pi^2 \epsilon} a_k^2 +
  \frac{1}{8\pi^2 \epsilon} \sum_{j=1}^{\nf} a_j^2 x_{jk}^2, 
\quad
  \mat{\delta Q}_{a_k}  = -\frac{1}{8\pi^2 \epsilon} \sum_{j=1}^{\nf}
  a_j^2 x_{jk} \left( \mat{\hat{a}}_j \wedge \mat{\hat{a}}_k \right) =
  \frac{1}{2} \left( \mat{\delta Z} \mat{\hat{a}}_k \right) \wedge
  \mat{\hat{a}}_k, 
\end{equation}
which fixes the renormalization of the Yukawa couplings $\{a_k\}$ and
versors $\{\mat{\hat{a}}_k\}$.  If the $\{\mat{\hat{a}}_k\}$ are
mutually orthogonal, $\mat{\delta Q}_{a_k} = 0$, and $\delta Z_{a_k} =
5 a_k^2/(16\pi^2 \epsilon)$ does not depend on $a_j$ with $j \neq k$.
In this case the fermion sector of the model reduces to $\nf$\ copies
of the case $\nf = 1$.  Similar simplifications occur if all
$\{\mat{\hat{a}}_k\}$ are collinear, or if they are divided into
mutually orthogonal subsets of collinear versors.  We discuss these
geometric issues in section \ref{sec:rgflow}, in connection with RG
equations.  The renormalization of the potential terms
$V_4(\mat{\phi})$ is discussed in appendix \ref{sec:appv4}.  Notice
that only the matrices $\mat{\delta Q}_{a_k}$ enter the counterterms
to $V_4$, but not the constants $\delta Z_{a_k}$.

The RG equations for the Yukawa couplings $\{a_k\}$ and versors
$\{\mat{\hat{a}}_k\}$ have the form,
\begin{equation}
  \label{eq:msyrg}  
  \mub \frac{d a_k}{d\mub} = \beta_{a_k},
  \qquad
  \mub \frac{d \mat{\hat{a}}_k}{d\mub} =
  \mat{\beta}_{\hat{a}_k}. 
\end{equation}
From (\ref{eq:yuk}), with the renormalization constants
(\ref{eq:msyct2}) and retaining terms through $\mathcal{O}(a_k^3)$, we
get,
\begin{equation}
  \label{eq:abet}
  \beta_{a_k} = -\frac{\epsilon}{2} a_k + \frac{5}{16\pi^2} a_k^3 +
  \frac{a_k}{8\pi^2}\sum_{k\neq j=1}^{\nf} x_{kj}^2 a_j^2.
\end{equation}
We see that if the coupling vectors are pairwise orthogonal, $x_{jk}
= 0$ if $j\neq k$, the RG equations for the $a_k$ decouple.  In the
general case the evolution is coupled, but it does not depend on the
sign of $x_{jk}$.  In particular, $\beta_{a_k} > 0$ for all $k$ at
$d=4$ so all Yukawa couplings grow with $\log(\mub)$ as expected.  For
the Yukawa versors we obtain,
\begin{equation}
  \label{eq:habet}
  \mat{\beta}_{\hat{a}_k} = \frac{1}{8\pi^2} \sum_{j=1}^{\nf} a_j^2
  x_{kj} \left( \mat{\hat{a}}_j - x_{kj} \mat{\hat{a}}_k \right).
\end{equation}
We discuss this equation in more detail below.

The RG evolution of the coupling constants in $V_4$ is needed in order
to obtain the RG equations for \mat{H} and \mat{M^2}, but only to
lowest order $\mub dg/d\mub = -\epsilon g + \mathcal{O}(g^3)$, and
analogously for $g_{ij}$ and $g_{ijkl}$.  From (\ref{eq:hr}) we obtain
a RG equation for \mat{H} with the renormalization constants
\mat{\delta Z_{H}}, \mat{\delta V} as given by (\ref{eq:heq}) with
$\mat{\delta Q} = 0$,
\begin{equation}
  \label{eq:rghr}
  \mub \frac{d\mat{H}}{d\mub} = \mat{\beta_H},
  \quad
  \mat{\beta_H} = -\frac{\epsilon}{2} \mat{H} + \epsilon
  \mu^{-\epsilon/2} \mat{\Delta H} + \frac{1}{8 \pi^2 \epsilon}
  \sum_{k=1}^{\nf} \left\{ \mat{a_k} \otimes
  \mat{a_k}, \mat{H} \right\}.
\end{equation}
\mat{\Delta H} gathers the divergent part of the contributions from
$V_4$ to the \mat{\phi}--$\chi$  vertex.  For $\ns=2$ it is given in
(\ref{eq:dhapp}). 

From (\ref{eq:mscounter}) we find,
\begin{equation}
  \label{eq:rgcounter}
  \mub \frac{dM^2_{ab}}{d\mub} =   - \left(
  \mat{\gamma_m} \right)_{ab} = \frac{1}{8\pi^2}
  \left( \mat{H}^2 \right)_{ab} + \frac{m_a^2 + m_b^2}{8 \pi^2}
  \sum_{k=1}^{\nf} (\mat{a}_k)_a (\mat{a}_k)_b + \epsilon
  \left(\Omega^{(3)}_{ab}\right)_{\mathrm{div.}},
\end{equation}
where \mat{\gamma_m}, which has mass dimension 1, is defined by this
equation.  The RG equations (\ref{eq:rghr}) and (\ref{eq:rgcounter})
are algebraically complicated, due in part to the contributions
to \mat{\beta_H}\ and \mat{\gamma_m}\ from diagrams with one insertion
of $V_4$.  We will only make the following simple remarks about these
equations.  If $\{\mat{\hat{a}}_k\}$ are eigenvectors of \mat{M^2}\ at
a given scale, they must lie on its orthogonal eigenspaces which,
since \mat{M^2}\ is regular, must be one-dimensional.  Therefore, any
two versors $\mat{\hat{a}}_k$ and $\mat{\hat{a}}_j$ must be either
collinear or orthogonal.  As discussed below, such configurations are
fixed points of the RG flow.  Thus, if we require $\mat{M^2}
\mat{\hat{a}}_k = m^2_{\mat{\hat{a}}_k} \mat{\hat{a}}_k$ for all $k$
at all scales, differentiating we obtain $\mat{\gamma_m}
\mat{\hat{a}}_k = \lambda_{\mat{\hat{a}}_k} \mat{\hat{a}}_k$ for some
eigenvalue $\lambda_{\mat{\hat{a}}_k}$, leading to $ [ \mat{M^2},
\mat{\gamma_m} ] = 0$ in \sy.  In short, $\{\mat{\hat{a}}_k\}$ can be
eigenvectors of \mat{M^2}\ at all scales only if \mat{M^2}\ and
\mat{\gamma_m}\ can be diagonalized simultaneously in \sy\ for all
$\mub$.  The converse is clearly valid.  If some $m_a^2$ are
degenerate, the versors $\mat{\hat{a}}_k$ can be eigenvectors of
\mat{M^2}\ without being at a RG fixed point.  Analogous
considerations hold for \mat{H}\ and \mat{\beta_H}.

The anomalous dimension of $\chi$ vanishes.  For the fermion fields,
with $\delta Z_{\psi_k}$ from (\ref{eq:mszpsi}) we find
$\gamma_{\psi_k} = -1/(32 \pi^2) a_k^2$.  From (\ref{eq:msctr}) we get
for the anomalous dimension of \mat{\phi},
\begin{equation}
  \label{eq:rgctr}
  \mat{\gamma} = - \frac{1}{\mat{Z}^{1/2}} \mub
  \frac{d}{d\mub} \mat{Z}^{1/2} =
  - \frac{1}{8 \pi^2} \sum_{k=1}^{\nf} a_k^2 \mat{\hat{a}}_k \otimes
  \mat{\hat{a}}_k. 
\end{equation}
\mat{\gamma}\ is symmetric, which is consistent with $\mat{\delta Q} =
0$.  Notice that at this order \mat{\phi} components in directions
orthogonal to \sy\ have vanishing anomalous dimension, and that only
$\beta_{a_k}$ contributes to \mat{\gamma}, but not
$\mat{\beta}_{\hat{a}_k}$.  This prompts us to look at the anomalous
dimension of $\mat{\hat{a}}_k \pdot \mat{\phi}$, which receives
contributions from both $\mat{\beta}_{\hat{a}_k}$ and $\mat{\gamma}$.
Those two contributions partially cancel each other and we get,
symbolically,
\begin{equation}
  \label{eq:rgctr1}
  \mub \frac{d}{d\mub}  \mat{\hat{a}}_k \pdot
  \mat{\phi} = \mat{\eta}_k \pdot \mat{\phi},
  \quad
  \mat{\eta}_k = -\frac{1}{8\pi^2} \sum_{j=1}^{\nf} a_j^2 x_{jk}^2
  \mat{\hat{a}}_k~.
\end{equation}
At this order the RG evolution of the components $\mat{\hat{a}}_k
\pdot \mat{\phi}$ of \mat{\phi} is decoupled from the evolution of
$\mat{\hat{a}}_j \pdot \mat{\phi}$ with $j \neq k$.

\subsection{RG flow in flavor space}
\label{sec:rgflow}

The RG equations for the Yukawa coupling constants $\{a_k\}$ as given
by (\ref{eq:msyrg}) and (\ref{eq:abet}) show that those couplings grow
monothonically with $\log (\mub)$.  Since $\beta_{a_k} > 0$ for all
$k$, there can be no fixed points.  In this section we consider the
qualitative analysis of the RG flow of the Yukawa coupling versors
$\{\mat{\hat{a}}_k\}$.  Specifically, we discuss its fixed points and
invariant manifolds, and their stability.

We find it convenient to describe the problem not in terms of the
$\{\mat{\hat{a}}_k\}$, but of its associated Gram matrix $\mat{x} \in
\mathbb{R}^{\nf \times \nf}$, $x_{ij} = \mat{\hat{a}}_i \pdot
\mat{\hat{a}}_j$.  \mat{x} is symmetric, with diagonal entries equal
to 1, and $\mathrm{rank} (\mat{x}) = \dy$.  We consider $x_{ij}$ with
$ 1 \leq i < j \leq \nf$\ as our basic variables.  If the quantity
$x_{ji}$ with $ i < j$ occurs in an equation, it is to be interpreted
as $ x_{ji} \equiv x_{ij}$ both in its value and as a variable.  In
the same sense, $ \beta_{x_{ji}} \equiv \beta_{x_{ij}} $ if $ i < j $,
and $ x_{ii} \equiv 1 $, $ \beta_{x_{ii}} \equiv 0 $.  There are then
$\binom{\nf}{2}$ independent variables.

From (\ref{eq:habet}) we have,
\begin{equation}
  \label{eq:xbet}
  \mub \frac{d}{d\mub} \mat{x} = \mat{\beta_x},
  \quad
  \beta_{x_{jk}} = \frac{1}{8\pi^2} \sum_{l=1}^{\nf} a_l^2 \left(
    x_{lk} (x_{lj}-x_{lk}x_{kj}) +  x_{lj} (x_{lk}-x_{lj}x_{jk})
  \rule{0ex}{2ex}\right). 
\end{equation}
By definition, $|x_{ij}| \leq 1$ for all $i,j$.  If $x_{ij} = \pm 1$
for some $i,j$, from (\ref{eq:xbet}) we see that $\beta_{x_{ij}}$
vanishes.  Thus, (\ref{eq:xbet}) is consistent with $|x_{ij}| \leq 1$.

We are interested in the fixed points of (\ref{eq:xbet}), zeros of
$\mat{\beta_x}$, which are independent of the values of $a_l > 0$,
$l=1,\ldots,\nf$. Those are given by all $\binom{\nf}{2}$--tuples
$\mat{\xo} = (\xo_{ij})$ with $i<j$, such that all $\xo_{ij} = 0, \pm
1$, and satisfy the geometric consistency condition,
\begin{equation}
  \label{eq:geo}
  \mathrm{if} \quad \xo_{ij} = \pm 1 \quad \mathrm{then} \quad
  \xo_{ij} \xo_{jk} = \xo_{ik} \quad
  \mathrm{for\;  all} \quad 1 \leq k \leq \nf.
\end{equation}
The meaning of this condition is apparent: given that
$\mat{\hat{a}}_i$ and $\mat{\hat{a}}_j$ are parallel, if
$\mat{\hat{a}}_k$ is parallel (resp.\ antiparallel, orthogonal) to
$\mat{\hat{a}}_j$, then it must be parallel (resp.\ antiparallel,
orthogonal) to $\mat{\hat{a}}_i$, and analogously if $\mat{\hat{a}}_i$
and $\mat{\hat{a}}_j$ are antiparallel.    

If \mat{\xo} is a fixed point with some components equal to -1, then
$\mat{\xo}^\prime$ with ${\xo}^{\prime}_{ij} = |\xo_{ij}|$ is also a
fixed point, as can be seen from (\ref{eq:geo}).  Since we can always
flip any $\mat{a}_k$ in $\Ll$ by reparametrizing the corresponding
fermion field as $\psi_k \rightarrow \gamma_5 \psi_k$, \mat{\xo} and
$\mat{\xo}^\prime$ are equivalent.  In particular, the linearized
\mat{\beta_x} function has the same eigenvalues and eigenvectors at
\mat{\xo} and $\mat{\xo}^\prime$, so we need consider only those fixed
points \mat{\xo} with $\xo_{ij} = 0$ or $1$ and satisfying
(\ref{eq:geo}).

The fixed points can be grouped in three types.  First, there is the
fixed point $\mat{\xo}$ with $ \xo_{ij} = 0 $ for all $i<j$,
corresponding to the Yukawa versors $\{ \mat{\hat{a}}_k \}$ being
pairwise orthogonal.  Clearly, in this case we must have $\dy = \rank
(\mat{\xo}) = \nf \leq \ns$. 
The linearized \mat{\beta_x} function is diagonal and
strictly positive definite,
\begin{equation}
  \label{eq:beta1}
  8\pi^2 \left. \frac{\partial \mat{\beta_{x}}_{ij}}{\partial x_{kl}}
  \right|_{\mat{\xo}} = \delta_{ik} \delta_{jl} \left( a_i^2 + a_j^2
  \right) > 0. 
\end{equation}
This fixed point is therefore IR stable.

The second type comprises fixed points with $x_{ij} = \pm 1$ for all
$i<j$, i.e., all $\mat{a}_k$ collinear, $\dy = 1$.  These are all
equivalent, in the sense defined above, to \mat{\xo} with $ \xo_{ij} =
1 $ for all $i<j$.  The linearized \mat{\beta_x} function is scalar
and strictly negative definite at these fixed points,
\begin{equation}
  \label{eq:beta2}
  8\pi^2 \left. \frac{\partial \mat{\beta_{x}}_{ij}}{\partial x_{kl}} 
  \right|_{\mat{\xo}} = -2 \delta_{ik} \delta_{jl} \sum_{m=1}^{\nf} 
  a_m^2 < 0, 
\end{equation}
which are then UV stable.

The third type of fixed points consists of those \mat{\xo} with not
all $\xo_{ij} = 0$ and not all $\xo_{ij} =\pm 1$.  In this case $1 <
\dy = \rank ( \mat{\xo} ) < \nf$, and only those \mat{\xo} with $\rank
( \mat{\xo} ) \leq \ns$ are possible.  These fixed points correspond
to the $\{ \mat{\hat{a}}_k \}$ lying along $r$ orthogonal directions
in flavor space, with $1< r < \nf$, so that at least two of them are
collinear, but some of them are orthogonal to each other.  As shown in
appendix \ref{sec:appbeta}, these are all saddle points.

The set of fixed points just described are precisely the points in
parameter space where $\Ll$ is approximately invariant under the
softly-broken discrete symmetries \zy, etc., discussed at the
beginning of this section.  An important particular case occurs when
at a fixed point of second or third type not only some coupling
vectors are collinear, but also their magnitudes are equal.  For
example, if $\mat{a}_i = \mat{a}_j$, $i \neq j$, and they are
orthogonal to all other coupling vectors, then from
(\ref{eq:msyrg})--(\ref{eq:habet}) we see that the equality is
maintained at all scales.  In that case, $\Ll$ is invariant under an
exact $SU(2)$ symmetry acting on the fermion fields $\psi_i$,
$\psi_j$, on top of the corresponding, generally approximate, discrete
symmetry.  If there are $1 \leq r < \nf$ groups of $n_i > 1$ equal
coupling vectors, with each group orthogonal to all other coupling
vectors, the fermion sector has an exact $SU(n_1) \times \dots \times
SU(n_r)$ invariance.
 
Going back to the general case, not all of the fixed points described
above can be reached for given values of $\nf$ and $\ns$, and for
given initial conditions for (\ref{eq:msyrg}) or (\ref{eq:xbet}).  For
instance, as mentioned above, the fixed point of first type is not
reachable if $\nf > \ns$.  More generally, we are led to ask how
linear dependence and independence of $\{ \mat{\hat{a}}_k\}$, and
$\dy$, evolve under the RG flow.

The Yukawa versors are linearly independent if and only if \mat{x} is
non-singular.  From (\ref{eq:xbet}), taking into account $x_{ij}
\equiv x_{ji}$ and $x_{ii} \equiv 1$, we obtain the RG equation for
$\det(\mat{x})$,
\begin{equation}
  \label{eq:detx}
  \mub \frac{d}{d\mub} \det(\mat{x}) = \gamma_d \det(\mat{x}),
  \qquad
  \gamma_d = -\frac{1}{4\pi^2} \sum_{i \neq j = 1}^{\nf} a_i^2
  x_{ij}^2 \leq 0.
\end{equation}
We see that if at a scale $\mub_\naught$ we have $\det(\mat{x}) = 0$
(resp.\ $>0$, $<0$), then $\det(\mat{x}) = 0$ (resp.\ $>0$, $<0$) at
all finite scales $\mub$.  Thus, linear (in)dependence of $\{
\mat{\hat{a}}_k \}$ is preserved by the RG flow (but not
asymptotically preserved, however, since at the UV fixed points $\dy =
1$).  Assume that at $\mub_\naught$ the set $\{ \mat{\hat{a}}_k \}$ is
linearly independent.  We denote by \cy\ the ``Yukawa simplex'' formed
by all convex combinations of $\{ \mat{\hat{a}}_k \}$, whose volume is
proportional to $\det (\mat{x})$.  From (\ref{eq:detx}),
$\det(\mat{x})|_{\mub} = (\mub/\mub_\naught)^{\gamma_d}
\det(\mat{x})|_{\mub_\naught}$ at one loop, so that \cy\ loses volume
as $\mub$ grows at a rate dictated by the anomalous dimension
$\gamma_d$ of $\det (\mat{x})$.  We notice also that the r.h.s.\ of
(\ref{eq:detx}) vanishes at all fixed points because $\det(\mat{x})$
vanishes in all cases, except for the first-type fixed point, at which
$\gamma_d =0$.

We can, further, consider the cofactors of order $k$, $C^{i_1\dots
  i_{\nf-k}}_{j_1\dots j_{\nf-k}}(\mat{x};\nf,k)$, $0 < k \leq
\nf$, $1 \leq i_1,\dots,i_{\nf-k}, j_1,\dots,j_{\nf-k} \leq \nf$,
defined up to a sign as the determinant of the $k \times k$
matrix obtained from \mat{x} by deleting rows $i_1,\dots,i_{N-k}$ and
columns $j_1,\dots,j_{N-k}$.  $C^{i_1\dots i_{\nf-k}}_{j_1\dots
  j_{\nf-k}}(\mat{x};\nf,k)$ can be written as
\begin{equation}
  \label{eq:cof}
  C^{i_1 \dots i_{N-k}}_{j_1 \dots j_{N-k}} (\mat{x}; N, k) =
  \frac{1}{k!} \sum_{m_1, 
  \dots , m_k = 1}^N \sum_{n_1, \dots , n_k = 1}^N
  \varepsilon_{m_1 \dots m_k i_1 \dots i_{N-k}}
  \varepsilon_{n_1 \dots n_k j_1 \dots j_{N-k}} x_{m_1 n_1} \cdots
  x_{m_k n_k}.
\end{equation}
Since \mat{x} is symmetric, in particular diagonalizable, we are
interested only in diagonal cofactors such as $C^{i_1\dots
  i_{\nf-k}}_{i_1\dots i_{\nf-k}}(\mat{x};\nf,k)$, which is the
determinant of the Gram matrix associated to $\{ \mat{\hat{a}}_k \}_{k
  \neq i_1,\dots,i_{\nf-k}}$.  In particular, $C(\mat{x};\nf,\nf) =
\det(\mat{x})$.  For a diagonalizable matrix such as \mat{x} it is
easily seen that $\rank (\mat{x}) = \dy$ (with $\dy \leq \nf$) if and
only if $C^{i_1\dots i_{\nf-k}}_{i_1\dots i_{\nf-k}}(\mat{x};\nf,k)=0$
for all $k > \dy$ and for all $i_1,$ $\dots, i_{\nf-k}$, but
$C^{i_1\dots i_{\nf-\dy}}_{i_1\dots i_{\nf-\dy}}(\mat{x};\nf,\dy) \neq
0$ for some set of indices $i_1\dots i_{\nf-\dy}$.

Using (\ref{eq:xbet}) and (\ref{eq:cof}) and taking into account that
$x_{ij} \equiv x_{ji}$ and $x_{ii} \equiv 1$, we obtain,
\begin{equation}
  \label{eq:cxbet}
  \begin{split}
    \mub \frac{d}{d\mub} C^{i_1\dots i_{\nf-k}}_{i_1\dots
      i_{\nf-k}} (\nf,k) & = \gamma_{k}^{i_1\dots i_{\nf-k}}
    C^{i_1\dots i_{\nf-k}}_{i_1\dots i_{\nf-k}}(\nf,k) -
    \frac{1}{4\pi^2}
    \sum_{m=1}^{\nf - k} a_{i_m}^2 C^{i_1\dots \widehat{i_m}\dots
      i_{\nf-k}}_{i_1\dots \widehat{i_m}\dots
      i_{\nf-k}}(\nf,k+1),\\
    \gamma_{k}^{i_1\dots i_{\nf-k}} & = -\frac{1}{4\pi^2}
      \sum_{m=1}^{\nf} a_m^2 \left( \sum_{\substack{n=1 \\ n \neq
      i_1,\dots, i_{\nf-k}}}^{\nf} x_{mn}^2 -1 \right),
  \end{split}
\end{equation}
where the argument $\mat{x}$ has been suppressed for brevity, and the
caret on an index indicates that index is to be omitted.  Setting
$k=\nf$ in (\ref{eq:cxbet}) we recover (\ref{eq:detx}).

Assume now that $\rank (\mat{x}) = \dy < \nf$
at some scale $\mub_\naught$.  Then $C(\mat{x};\nf,\nf) = 0$ at all
scales by (\ref{eq:detx}) and, from (\ref{eq:cxbet}), the
$(\nf-1)$-cofactor satisfies,
\begin{equation}
  \label{eq:cxn-1}
\mub \frac{d}{d\mub} C^{i}_{i} (\mat{x}; \nf,\nf-1)  = \gamma_{\nf-1}^{i}
    C^{i}_{i}(\mat{x}; \nf,\nf-1).  
\end{equation}
If $\dy = \nf -1$, then at $\mub_\naught$ we must have $C^{j}_{j}
(\mat{x}; \nf,\nf-1) \neq 0$ for some $j$.  From (\ref{eq:cxn-1}) we
see that in that case $C^{j}_{j} (\mat{x}; \nf,\nf-1) \neq 0$ at all
finite scales $\mub$.  If $\dy < \nf -1$, then $C^{j}_{j} (\mat{x};
\nf,\nf-1) = 0$ at $\mub_\naught$ for all $j$, and by (\ref{eq:cxn-1})
they vanish at all finite scales.  Iterating this argument, we
conclude that if at some scale $\mub_\naught$ we have $C^{i_1\dots
  i_{\nf-k}}_{i_1\dots i_{\nf-k}} (\mat{x};\nf,k) =0$ for all
$i_1,\dots,i_{\nf-k}$ and all $k>\dy$, and $C^{i_1\dots
  i_{\nf-\dy}}_{i_1\dots i_{\nf-\dy}} (\mat{x};\nf,\dy) \neq 0$ for
some set of indices $i_1,\dots,i_{\nf-\dy}$, then that situation
persists at all finite scales $\mub$.  Therefore, $\rank (\mat{x}) =
\dy$ is RG invariant (though not asymptotically invariant).

For $k<\nf$, $\gamma_k^{i_1\dots i_{\nf-k}}$ can be positive, negative
or zero, depending on the values of $a_k$ and $x_{ij}$.  In a small
neighborhood of an UV fixed point, $x_{ij}^2$ will be close enough to
1 for all $i,j$, so that $\gamma_k^{i_1\dots i_{\nf-k}} < 0$ for $1< k
\leq \nf$.  In that situation, (\ref{eq:cxbet}) with $k=\dy$ describes
how \cy\ loses \dy-dimensional volume as $\mub$ grows.  Needless to
say, we are implicitly assuming that even though $\mat{x}$ is near an
UV fixed point, possibly as a result of a choice of initial
conditions, the couplings $a_k$ are small enough for perturbation
theory to be valid.

\section{An \mat{SU(N)}-symmetric model with SSB}
\label{sec:2model}

In this section we turn to an $SU(N)$ symmetric model with scalar
fields in the adjoint representation, $\vect{\phi} = (\phi_1, \dots,
\phi_{\H{N}})$, and fermion fields in the fundamental representation,
$\mat{\psi} = (\psi_1, \dots, \psi_N)$.  We denote $\H{n}\equiv n^2-1$
for brevity, $\vect{\lambda} = (\mat{\lambda}_1, \dots,
\mat{\lambda}_{\H{N}})$ for the vector of Gell-Mann matrices as
defined in appendix \ref{sec:appsun}, and $\mat{\phi} = \vect{\phi}
\pdot \vect{\lambda}$.  The classical Lagrangian in the broken
symmetry phase of the model has the form,
\begin{equation}
  \label{eq:sunlag}
  \Ll = -\frac{1}{2} \vect{\phi} \pdot (\Box - \nu^2)
  \vect{\phi} - V_4(\vect{\phi},N) + \overline{\mat{\psi}} i
  \dirac \mat{\psi} - g \overline{\mat{\psi}} \mat{\phi} \mat{\psi},
\end{equation}
with $\nu^2 > 0$ and $V_4(\vect{\phi},N)$ the most general
$SU(N)$ invariant quartic homogeneous polynomial in \vect{\phi},
as given below.  $\Ll$ is invariant under the $U(1)$ symmetry related
to fermion number conservation, the \zz\ chiral symmetry
$\vect{\phi} \rightarrow -\vect{\phi}$, $\mat{\psi}
\rightarrow \gamma_5 \mat{\psi}$, and the $SU(N)$ transformations,
\begin{equation}
  \label{eq:suntrans}
  \mat{\psi} \rightarrow e^{-i\vect{\theta}\cdot\vect{\lambda}/2}
  \mat{\psi},
  \quad
  \phi_a \rightarrow \sum_{b=1}^{\H{N}} R_{ab}(\vect{\theta}) \phi_b,
  \quad
  R_{ab}(\vect{\theta}) = \frac{1}{2}
  \Tr{e^{i\vect{\theta}\cdot\vect{\lambda}/2} \mat{\lambda}_a
  e^{-i\vect{\theta}\cdot\vect{\lambda}/2} \mat{\lambda}_b}.
\end{equation}
The $SU(N)$ currents are given by $j^{\mu}_c = \overline{\mat{\psi}}
\gamma^\mu \mat{\lambda}_c/2 \mat{\psi} - \sum_{ab} f_{cab}
\partial^\mu \phi_a \phi_b$, with $f_{cab}$ the $SU(N)$ structure
constants.  Both \zz\ and $SU(N)$ symmetries are broken if $\langle
\vect{\phi} \rangle \neq 0$.

We can build two quartic $SU(N)$ singlets out of \mat{\phi},
$(\sum_{ij} \phi_{ij}\phi_{ji})^2$ and $\sum_{ijkl}
\phi_{ij}\phi_{jk}\phi_{kl}\phi_{li}$, giving rise to two terms in
$V_4$.  Given two vectors \vect{\alpha}, \vect{\beta}\ in the adjoint
representation, we introduce the notation $(\vect{\alpha} \vee
\vect{\beta})_a = \sum_{b,c=1}^{\H{N}} d_{abc} \alpha_b \beta_c$, with
$d_{abc}$ the anticommutator constants for $\vect{\lambda}$ matrices.
With these notations, $V_4$ can be variously written as,
\begin{equation}
  \label{eq:sunv4}
  \begin{split}
    V_4(\vect{\phi},N) & = \frac{\lambda}{4} \left( \vect{\phi}^2
    \right)^2 - \frac{\lambda^\prime}{4} \left( \vect{\phi} \vee
      \vect{\phi} \right)^2 = \frac{\lambda}{4} \left(
      \sum_{a=1}^{\H{N}} \phi_a^2 \right)^2 - \frac{\lambda^\prime}{4}
    \sum_{abcdh=1}^{\H{N}} d_{abh}
    d_{cdh} \phi_{a} \phi_{b} \phi_{c} \phi_{d}\\
    & = \frac{1}{8} \left( \frac{\lambda}{2} +
      \frac{\lambda^\prime}{N} \right)
    \left(\Tr{\mat{\phi}^2}\right)^2 - \frac{\lambda^\prime}{8}
    \Tr{\mat{\phi}^4} = V_4(\mat{\phi},N).
  \end{split}
\end{equation}
The coupling constants $\lambda$, $\lambda^\prime$ are defined to be
positive.  The choice of sign for the term in $\lambda^\prime$ in
(\ref{eq:sunv4}) will be explained shortly.  It is easily seen,
however, that for appropriate ranges of positive $\lambda$ and
$\lambda^\prime$, $V_4$ is bounded below.  Given any Hermitian matrix
\mat{A}, not necessarily traceless, we can evaluate $V_4(\mat{A},N)$
in a flavor basis such that $\mat{A}=\diag(\alpha_1,\dots,\alpha_N)$,
with $\alpha_j$ real,
\begin{equation*}
  V_4(\mat{A},N) = \frac{1}{8} \left( \frac{\lambda}{2} -
  \frac{N-1}{N} \lambda^\prime  \right) \sum_{j=1}^N \alpha_j^4 +
  \frac{1}{8} \left( \frac{\lambda}{2} + 
  \frac{\lambda^\prime}{N}  \right) \sum_{i \neq j=1}^N \alpha_i^2
  \alpha_j^2.  
\end{equation*}
We see that if we choose $\lambda^\prime/\lambda < N/(2(N-1))$, then
$V_4(\mat{A},N) \geq 0$ for all Hermitian \mat{A}.  Setting
$\lambda^\prime/\lambda < 1/2$ we ensure that $V_4$ is bounded below,
and in fact positive, for all $N \geq 2$.

We denote the local minima of the tree-level potential
$\V(\vect{\phi},N) = V_4(\vect{\phi},N) - \nu^2/2\, \vect{\phi}^2$ by
$\mat{v}_{(0)} = \vect{v}_{(0)} \pdot \vect{\lambda} = v_{(0)}
\mat{\hat{v}}_{(0)} \pdot \vect{\lambda}$.  Due to $SU(N)$ invariance
the manifold of minima of $\V(\mat{\phi},N)$ consists of a disjoint
union of $SU(N)$ orbits, each of which contains several diagonal
matrices with the same eigenvalues, related to each other by $SU(N)$
transformations which are reorderings of the flavor basis.  Each one
of these orbits constitutes an $SU(N)$ equivalence class of minima of
\V, or ``modulus'' for short.  We can choose a diagonal matrix
$\mat{v}^\prime_{(0)} = v_{(0)}^\prime \sum_{n=2}^N
(\mat{\hat{v}}_{(0)}^\prime)_{\H{n}} \mat{\lambda}_{\H{n}}$, with
$\mat{\hat{v}}_{(0)}^\prime \pdot \mat{\hat{v}}_{(0)}^\prime =
\sum_{n=2}^N (\mat{\hat{v}}_{(0)}^\prime)_{\H{n}}^2 = 1$, out of each
orbit as representative of the corresponding equivalence class.  Thus,
we can parametrize classical moduli space as a submanifold of the
algebra of diagonal matrices of $su(N)$ generated by $\{
\mat{\lambda}_{\H{n}}\}$, $n=2,\dots,N$.

For $N=2$, $d^{abc} = 0$ and $\V(\vect{\phi},2)$ is isotropic in
flavor space.  Its minima are all vectors $\vect{v}_{(0)}$ with
$v_{(0)} = \nu/\sqrt{\lambda}$, all of them lying in the $SU(2)$ orbit
of $\mat{v}_{(0)}^\prime = v_{(0)} \mat{\lambda}_3$.  The space of
moduli consists in this case of a single point.  There are two
Goldstone bosons and a massive scalar with tree-level mass $\sqrt{2}
\nu$, and the two fermion fields are mass-degenerate.

The case $N=3$ is special. Using the fact that $\mat{\phi}$ is a root
of its characteristic polynomial and that $\tr(\mat{\phi}) = 0$, we
find that $\tr(\mat{\phi}^4) = 1/2 (\tr(\mat{\phi}^2))^2$
\cite{cole}.  In this case also $V_4 (\vect{\phi},3)$ is
isotropic, $V_4 (\mat{\phi},3) = 1/16 (\lambda - \lambda^\prime/3) (
\Tr{\mat{\phi}^2} )^2.$ (We notice, incidentally, that we can take
$\lambda^\prime/\lambda$ as large as 3 without losing positivity of
$V_4(\mat{\phi},3)$.)  From this expression, and (\ref{eq:sunlag}), we
see that at tree level the scalar sector of the model is invariant
under the action of the fundamental representation of the group
$SO(8)$.  The classical minima of the potential are the vectors
$\vect{v}_{(0)}$ with $v_{(0)} = \nu /
\sqrt{\lambda-\lambda^\prime/3}$.  Once a direction for
$\vect{v}_{(0)}$ is chosen its stationary group is $SO(7)$, so at
tree level we have seven massless and one massive scalar field with
mass $\sqrt{2}\nu$.  Clearly, $SO(8)$ is not a symmetry of the model,
since it is explicitly broken by the fermion sector.  In particular
\textsc{vev}s $\mat{v}_{(0)}$ with different eigenvalues, which
correspond to different moduli, lead to different fermion mass-spectra
at tree level.

Moduli space can be parametrized by the circle
$\{\mat{v}_{(0)}^\prime(\theta) = v_{(0)} (\cos\theta \mat{\lambda}_3
+ \sin\theta \mat{\lambda}_8)\}$.  Since $\mat{v}_{(0)}^\prime
(\theta)$ and $\mat{v}_{(0)}^\prime (\theta + 2 \pi/3)$ differ only by
a permutation of their diagonal elements, if we want to factor the
action of the permutation subgroup $\mathbb{Z}_3$ of $SU(3)$ we must
restrict the range of $\theta$ to $-\pi/3 < \theta < \pi/3$.  All
$\mat{v}_{(0)}^\prime(\theta)$ with $\theta = \pi/2 + n\pi/3$, in
particular $\mat{v}_{(0)}^\prime(\pi/2)=\mat{\lambda}_8$, have two
equal eigenvalues.  Thus, their stationary subgroups are isomorphic to
$SU(2)\times U(1)$.  In this case there are four Goldstone bosons,
three scalar fields acquire mass radiatively, and tree-level fermion
mass ratios are 1:1:2.  On the other hand,
$\mat{v}_{(0)}^\prime(\theta)$ with $\theta \neq \pi/2 + n\pi/3$ have
three different eigenvalues, and a stationary subgroup isomorphic to
$U(1)\times U(1)$.  There are six Goldstone bosons in this case, with
one scalar acquiring mass radiatively.  All vacua are degenerate at
tree level.

For $N>3$, $V_4(\mat{\phi},N)$ is not isotropic in flavor space.  Like
in the previous case, we can parametrize moduli space by a set of
minima $\mat{v}_{(0)}^\prime$ of $V_4$ within the subalgebra of
diagonal matrices of $su(N)$.  Unlike the case $N=3$, for $N>3$ moduli
space consists of a finite set of isolated points.  Rather than trying
to map all of moduli space for all $N>3$, in this paper we restrict
ourselves to pointing out that for all $N \geq 2$ there is a local
minimum of the potential of the form $\mat{v}_{(0)} = v_{(0)}
\mat{\lambda}_{\H{N}}$.  This minimum has the largest possible
stationary group, $SU(N-1) \times U(1)$, leading to the fewest
possible Goldstone bosons, $2N-2$.

Evaluating the extremum conditions $\partial
\V(\vect{\phi},N)/\partial \phi_t = 0$, $t=1,\dots,\H{N}$, for the
tree-level potential at $\mat{v}_{(0)} = v_{(0)} \mat{\lambda}_{\H{N}}$
(i.e., $(\vect{v}_{(0)})_a = v_{(0)} \delta_{a \H{N}})$, we obtain
\begin{equation}
  \label{eq:v0}
  v_{(0)} = \frac{\nu}{\sqrt{\lambda - D^2
      \lambda^\prime}},  \qquad
  D\equiv d_{\H{N}\H{N}\H{N}} = -(N-2) \sqrt{\frac{2}{N(N-1)}},
\end{equation}
where use was made of the properties of the constants $d_{abc}$ with
one or more indices equal to $\H{N}$, as given in appendix
\ref{sec:appsun}.  With the assumption $\lambda^\prime < \lambda/2$,
the argument of the square root is positive for $N \geq 3$.  At this
extremum second derivatives $\partial^2 \V/\partial \phi_s \partial
\phi_t$ with $s \neq t$ vanish.  For $s=t$ we find,
\begin{equation}
    \left. \frac{\partial^2 \V}{\partial \phi_{\H{N}}^2}
    \right|_{\mat{\phi} = v_{(0)} \mat{\lambda}_{\H{N}}} = 2 \nu^2 >
    0,
    \quad \mathrm{and} \quad
    \left. \frac{\partial^2 \V}{\partial \phi_{\H{n}}^2}
    \right|_{\mat{\phi} = v_{(0)} \mat{\lambda}_{\H{N}}}  = 2
    \lambda^\prime v_{(0)}^2 \frac{N-3}{N-1}
    \quad \mathrm{if} \quad
    2 \leq n < N.  \label{eq:jac}
\end{equation}
The r.h.s.\ in the second eq.\ in (\ref{eq:jac}) vanishes if $N=3$,
and it is positive for $N>3$ as long as $\lambda^\prime >0$, which
explains our choice of sign in (\ref{eq:sunv4}).  Due to $SU(N)$
invariance, there is no need to evaluate the second derivatives in the
directions of the non-diagonal generators: $\partial^2 \V/\partial
\phi_{t}^2$ with $1 \leq t \leq \H{N-1}-1$ and $t \neq \H{n}$ leads to
the same result as in (\ref{eq:jac}) with $t=\H{n}$, $n<N$.  If we set
$\H{N-1} < t < \H{N}$, we get $\partial^2\V/\partial \phi_t^2 = 0$ at
$\mat{v}_{(0)}$, as required by Goldstone's theorem.  Thus,
$\mat{v}_{(0)} = v_{(0)} \mat{\lambda}_{\H{N}}$ is a minimum of the
tree-level potential for $N \geq 3$ if $\lambda^\prime > 0$.  If
$\lambda^\prime < 0$ and $N>3$ it is a saddle point.  We may also ask
whether there are other minima $ \propto \mat{\lambda}_{\H{n}}$, with
$2\leq n < N$.  This occurs only for $N=3$.  For $N>3$, there is an
extremum of the potential only in the direction of $\mat{\lambda}_3$
but numerically it turns out to be a saddle point for $N=4$ and 5.
All other minima in the subalgebra of diagonal matrices of $su(N)$
must be linear combinations of more than one $\mat{\lambda}_{\H{n}}$.

Shifting the field $\vect{\phi} = \vect{\varphi} + \vect{v}_{(0)}$,
and substituting in the classical Lagrangian (\ref{eq:sunlag}), we
obtain the tree-level masses for the scalar fields
\begin{equation}
  \label{eq:smass}
  m_a^2 = 2 \lambda^\prime v_{(0)}^2 \frac{N-3}{N-1} \quad
  \mathrm{if} \quad 1 \leq a \leq \H{N-1},
  \quad
  m_a^2 = 0 \quad \mathrm{if} \quad (N-1)^2 \leq a < \H{N},
  \quad
  m^2_{\H{N}} = 2 \nu^2.
\end{equation}
For $N=3$, as expected, $m_a^2 =0$ for $a < 7$,  $\varphi_{1,2,3}$
acquiring mass at one loop.  For the fermion sector we have the mass
matrix $\mat{M_\psi} = g v_{(0)} \mat{\lambda}_{\H{N}}$ or, more
explicitly,
\begin{equation}
  \label{eq:fmass}
  m_{\psi_i} = m_\psi \quad \mathrm{if} \quad 1\leq i \leq N-1,
  \quad
  m_{\psi_N} = -(N-1) m_\psi,
  \quad
  m_\psi \equiv g v_{(0)} \sqrt{\frac{2}{N(N-1)}}.
\end{equation}
Notice that, at tree level, $\sum_{i=1}^N m_{\psi_i} = 0$ and
$\sum_{i=1}^N m_{\psi_i}^2 = 2 g^2 v_{(0)}^2$.  For the purpose of the
perturbative expansion we consider $\lambda^\prime$ to be of the same
order of magnitude as $\lambda$, and $g=\mathcal{O}(\lambda^{1/2})$,
so that $m_\psi=\mathcal{O}(1)$.

In the quantum Lagrangian we introduce renormalization constants for
the parameters, $\nu_\naught^2 = \nu^2 + \delta\nu^2$,
$\lambda_\naught = \mu^\epsilon (\lambda + \delta \lambda)$,
$\lambda_\naught^\prime = \mu^\epsilon (\lambda^\prime + \delta
\lambda^\prime)$ and $g_\naught = \mu^{\epsilon/2} Z_g g$.  We
restrict ourselves to \msb\ scheme.  Thus, wave function
renormalization matrices reduce to the scalars $\delta Z$ for
$\vect{\phi}$ and $\delta Z_\psi$ for $\mat{\psi}$, due to $SU(N)$
symmetry.  We write \Ll\ in terms of shifted fields $\vect{\phi} =
\vect{\varphi} + \mu^{-\epsilon/2} \vect{v}$, where $\mu$ is the \ms\ 
mass scale, $\epsilon = 4 - d$, and the \textsc{vev} is given by
$\vect{v} = \vect{v}_{(0)} + \vect{v}_{(1)}$ with $\vect{v}_{(0)}$ the
tree-level \textsc{vev}, $\vect{v}_{(0)}\pdot \vect{\lambda} = v_{(0)}
\mat{\lambda}_{\H{N}}$, and $\vect{v}_{(1)} =
\mathcal{O}(\lambda^{1/2})$ to be determined by the requirement that
$\langle 0 | \vect{\varphi} | 0 \rangle = 0$ at one loop.

With these definitions, and retaining only terms contributing at
one-loop level, \Ll\ is written as,
\begin{align}
  \label{eq:sunqlag}
  \Ll &= \Ll_f + \Ll_1 + \Ll_2 + \Ll_\mathrm{int} + \Ll_\mathrm{ct} \\
  \Ll_f &= -\frac{1}{2} \vect{\varphi}\pdot \Box \vect{\varphi} +
  \frac{1}{2} \vect{\varphi}\pdot \mat{M} \vect{\varphi} +
  \overline{\mat{\psi}} i \dirac \mat{\psi} - \overline{\mat{\psi}}
  \mat{M_\psi} \mat{\psi} \nonumber\\
  \Ll_1 &= \mu^{-\frac{\epsilon}{2}}\nu^2 \vect{v}_{(1)}\pdot
  \vect{\varphi} - \mu^{-\frac{\epsilon}{2}} \lambda v_{(0)}^2
  \vect{v}_{(1)}\pdot \vect{\varphi} - 2 \mu^{-\frac{\epsilon}{2}}
  \lambda \left(\vect{v}_{(0)}\pdot
    \vect{v}_{(1)}\right)\left(\vect{v}_{(0)}\pdot
    \vect{\varphi}\right) \nonumber\\
  &\quad + \mu^{-\frac{\epsilon}{2}} \lambda^\prime
  \left(\vect{v}_{(0)} \vee
    \vect{v}_{(0)}\right) \pdot
  \left(\vect{v}_{(1)} \vee \vect{\varphi}\right) + 2
  \mu^{-\frac{\epsilon}{2}} \lambda^\prime \left(\vect{v}_{(0)}
    \vee \vect{v}_{(1)}\right) \pdot
  \left(\vect{v}_{(0)}
    \vee \vect{\varphi}\right) \nonumber\\
  \Ll_2 &= -\lambda \left(\vect{v}_{(0)}\pdot
    \vect{v}_{(1)}\right) \vect{\varphi}^2 - 2 \lambda
  \left(\vect{v}_{(0)}\pdot
    \vect{\varphi}\right)\left(\vect{v}_{(1)}\pdot
    \vect{\varphi}\right) + \lambda^\prime
  \left(\vect{v}_{(0)} \vee
    \vect{v}_{(1)}\right) \pdot \left(\vect{\varphi} \vee
    \vect{\varphi}\right) \nonumber\\
  &\quad + 2 \lambda^\prime \left(\vect{v}_{(0)} \vee
    \vect{\varphi}\right) \pdot \left(\vect{v}_{(1)} \vee
    \vect{\varphi}\right) - g \overline{\mat{\psi}}
  \vect{v}_{(1)}\pdot
  \vect{\lambda} \mat{\psi} \nonumber\\
  \Ll_\mathrm{int} &= -\mu^{\frac{\epsilon}{2}} \lambda
  \left(\vect{v}_{(0)}\pdot \vect{\varphi}\right)
  \vect{\varphi}^2 + \mu^{\frac{\epsilon}{2}} \lambda^\prime
  \left(\vect{v}_{(0)} \vee \vect{\varphi}\right) \pdot
  \left(\vect{\varphi} \vee \vect{\varphi}\right) - \mu^\epsilon
  \frac{\lambda}{4} \left(\vect{\varphi}^2\right)^2 + \mu^\epsilon
  \frac{\lambda^\prime}{4} \left(\vect{\varphi} \vee
    \vect{\varphi}\right)^2 - g \mu^{\frac{\epsilon}{2}}
  \overline{\mat{\psi}} 
  \vect{\varphi} \pdot \vect{\lambda}
  \mat{\psi} \nonumber\\
  \Ll_\mathrm{ct} &= -\frac{1}{2} \delta Z \vect{\varphi}\pdot \Box
  \vect{\varphi} + \delta Z_\psi \overline{\mat{\psi}} i \dirac
  \mat{\psi} - g \Delta Z_g \overline{\mat{\psi}}
  \vect{v}_{(0)}\pdot \vect{\lambda} \mat{\psi} - g
  \mu^{\frac{\epsilon}{2}} \Delta Z_g \overline{\mat{\psi}}
  \vect{\varphi} 
  \pdot \vect{\lambda} \mat{\psi} \nonumber\\
  &\quad + \mu^{-\frac{\epsilon}{2}} \left(\Delta \nu^2 - v_{(0)}^2 
    \Delta \lambda\right) \vect{v}_{(0)} \pdot
  \vect{\varphi} + \mu^{-\frac{\epsilon}{2}} \Delta \lambda^\prime 
  \left(\vect{v}_{(0)} \vee
    \vect{v}_{(0)}\right) \pdot
  \left(\vect{v}_{(0)} \vee \vect{\varphi}\right) +
  \frac{1}{2} \left(\Delta \nu^2 - v_{(0)}^2 \Delta
    \lambda\right) \vect{\varphi}^2  \nonumber\\
  &\quad - \Delta \lambda \left(\vect{v}_{(0)} \pdot
    \vect{\varphi}\right)^2 + \frac{\Delta \lambda^\prime}{2}
  \left(\vect{v}_{(0)} \vee
    \vect{v}_{(0)}\right) \pdot \left(\vect{\varphi} \vee
    \vect{\varphi}\right) + \Delta \lambda^\prime
  \left(\vect{v}_{(0)} \vee \vect{\varphi}\right)^2 -
  \mu^{\frac{\epsilon}{2}} \Delta \lambda \left(\vect{v}_{(0)} \pdot 
  \vect{\varphi}\right) \vect{\varphi}^2  \nonumber\\
  &\quad + \mu^{\frac{\epsilon}{2}} \Delta \lambda^\prime
  \left(\vect{v}_{(0)} \vee \vect{\varphi}\right) \pdot
  \left(\vect{\varphi} \vee \vect{\varphi}\right) -\mu^\epsilon
  \frac{\Delta \lambda}{4} \left(\vect{\varphi}^2\right)^2 +
  \mu^\epsilon \frac{\Delta \lambda^\prime}{4}
  \left(\vect{\varphi}\vee \vect{\varphi}\right)^2, \nonumber
\end{align}  
where \mat{M_\psi}\ is given by (\ref{eq:fmass}), $(\vect{v}_{(0)})_a
= v_{(0)} \delta_{a\H{N}}$ with $v_{(0)}$ from (\ref{eq:v0}),
$\vect{\varphi}\pdot \mat{M} \vect{\varphi} = \sum_{a=1}^{\H{N}} m_a^2
\varphi_a^2$ with $m_a^2$ from (\ref{eq:smass}), and
\begin{equation}
  \label{eq:sunct }
  \Delta \lambda = \delta \lambda + 2 \lambda \delta Z,
  \quad
  \Delta \lambda^\prime = \delta \lambda^\prime + 2 \lambda^\prime
  \delta Z,
  \quad
  \Delta \nu^2 = \delta \nu^2 + \nu^2 \delta Z,
  \quad
  \Delta Z_g = \delta Z_g + \delta Z_\psi + \frac{1}{2} \delta Z.
\end{equation}
The Lagrangian $\Ll_1 = \mathcal{O}(\lambda^{1/2})$, and $\Ll_2 =
\mathcal{O}(\lambda)$.

The condition that the finite part of $\langle 0 | \vect{\varphi}(0) |
0 \rangle$ must vanish at one loop leads to $(\vect{v}_{(1)})_a = 0$
for $a < \H{N}$ and,
\begin{align}
  \label{eq:v1}
    2 \nu^2 (\vect{v}_{(1)})_{\H{N}} &= -\frac{1}{16\pi^2} m_1^2
    a_0\left(m_1^2\right) v_{(0)} N (N-2) \left( \lambda +
      \lambda^\prime \frac{2 (N-4)}{N(N-1)} \right) -
    \frac{3}{16\pi^2} v_{(0)} m_{\H{N}}^2 a_0\left(m_{\H{N}}^2\right)
    \left(
      \lambda \rule{0pt}{3ex}\right. \nonumber\\
    &\left. - \lambda^\prime \frac{2 (N-2)^2}{N (N-1)}\right) +
    \frac{g^4 v_{(0)}^3}{\pi^2} \frac{1}{N^2(N-1)}
    a_0\left(m_\psi^2\right) + \frac{g^4 v_{(0)}^3}{\pi^2}
    \frac{(N-1)^2}{N^2} a_0\left(m_{\psi_N}^2\right).
\end{align}
At $N=3$, $\vect{v}_{(1)}$ depends on $\lambda$ and $\lambda^\prime$
only through $\lambda - \lambda^\prime/3$, as it should.  The
requirement that the divergent piece of the $\vect{\varphi}$ 1-point
function vanishes yields the relation
\begin{multline}
  \label{eq:ct1}
  \Delta \nu^2 - v_{(0)}^2 \Delta \lambda + v_{(0)}^2
  \frac{2(N-2)^2}{N(N-1)} \Delta \lambda^\prime =  
  -\frac{3}{4\pi^2 \epsilon} \nu^2 \left(\lambda -
    \frac{2(N-2)^2}{N(N-1)} \lambda^\prime \right) \\
  - \frac{1}{4\pi^2 \epsilon} v_{(0)}^2
  \frac{(N-3)(N-2)N}{N-1} \lambda^\prime \left( \lambda +
  \frac{2(N-4)}{N(N-1)}\lambda^\prime\right) + \frac{2}{\pi^2 \epsilon}
  \frac{N^2-3N+3}{N(N-1)} v_{(0)}^2 g^4. 
\end{multline}
Together with two other similar equations from the 2-point function,
(\ref{eq:ct1}) determines the counterterms on the l.h.s..

The self-energy function for the $\vect{\varphi}$ field must be
symmetric in flavor space, due to $CPT$ invariance, and invariant
under the $SU(N-1)\times U(1)$ symmetry of $\Ll$ and $\vect{v}$.  Those
two requirements together imply that it must be diagonal, of the form
$\diag(a, \ldots, a,b,\ldots,b,c)$ where the first block is
$\H{N-1}\times \H{N-1}$, the second $(2N-2) \times (2N-2)$ and the
third $1\times 1$.  More generally, the triviality of mixing in this
model holds for any pattern of symmetry breaking.  Any \textsc{vev} of
the scalar field has a stationary group containing a subgroup of
$SU(N)$ isomorphic to $U(1)^{N-1}$.  This symmetry, together with the
fact that the scalar field self-energy must be symmetric in its flavor
indices due to $CPT$ invariance, is enough to make the self-energy
function diagonal in the basis in which the \textsc{vev} is.

At one loop the self-energy function can be written as,
\begin{align}
  \label{eq:2ptfn}
  \Pi_{ab}(p^2) & = \Pi_{ab}^{(0)}(p^2) + \Pi_{ab}^{(1)}(p^2) +
  \sum_{j=1}^4   \Pi_{ab}^{(2.j)}(p^2),\\
  \Pi_{ab}^{(0)}(p^2) &= \delta_{ab} \left( 2 \lambda \vect{v}_{(0)}
    \pdot \vect{v}_{(1)} - \Delta \nu^2 + \Delta \lambda v_{(0)}^2
  \right) + \delta_{a\H{N}} \delta_{b\H{N}} 2 \left(2 \lambda
    \vect{v}_{(0)} \pdot \vect{v}_{(1)} + \Delta \lambda
    v_{(0)}^2 \right) \nonumber\\
  &\quad -2 \left( \vect{v}_{(0)} \pdot \vect{v}_{(1)} \lambda^\prime
    + \frac{v_{(0)}^2}{2} \Delta \lambda^\prime \right) \left(
    \rule{0pt}{2ex}
    d_{\H{N}\H{N}\H{N}} + 2 d_{\H{N}aa}\right) d_{\H{N}aa} \delta_{ab} -p^2
  \delta_{ab} \delta Z,
  \nonumber\\
  \Pi_{ab}^{(1)}(p^2) &=   \delta_{ab} \lambda \left( \sum_{d=1}^{\H{N}}
  A_0(m_d^2) + 2 A_0(m_a^2) \right) - \lambda^\prime
  \sum_{l,n=1}^{\H{N}} \left( d_{lnn} d_{lab} + 2 d_{lna} d_{lnb}
  \rule{0pt}{2ex} \right) A_0(m_n^2), \nonumber \\
%
%
  \Pi_{ab}^{(2.1)}(p^2) &=  2 v_{(0)}^2 \lambda^2 \left\{ \delta_{ab}
    2 B_0(p^2,m_{\H{N}}^2,m_a^2) 
    + \delta_{a\H{N}} \delta_{b\H{N}} \left(
      \sum_{c=1}^{\H{N}} B_0(p^2,m_c^2,m_c^2)+ \right.\right. \nonumber\\
    &\quad \left.\left.  + 4   B_0(p^2,m_{\H{N}}^2,m_{\H{N}}^2)
    \rule{0pt}{3.5ex}\right)\right\}, \nonumber\\
  \Pi_{ab}^{(2.2)}(p^2) &= 2 v_{(0)}^2 \lambda^{\prime 2}
  \sum_{c,d=1}^{\H{N}} \left\{ d_{\H{N}aa}d_{\H{N}bb} + 2 d_{\H{N}aa}
    d_{\H{N}dd} + 2 d_{\H{N}cc} d_{\H{N}bb} + 2 (d_{\H{N}cc})^2
    \right.\nonumber\\
    &\quad \left. + 2 d_{\H{N}cc} d_{\H{N}dd} \rule{0pt}{2ex}\right\}
    d_{acd} d_{bcd} B_0(p^2,m_c^2,m_d^2), \nonumber\\
  \Pi_{ab}^{(2.3)}(p^2)  &= -2 v_{(0)}^2 \lambda\lambda^{\prime}
  \left\{ \delta_{a\H{N}} \sum_{c=1}^{\H{N}} d_{bcc} (d_{\H{N}bb}+2
  d_{\H{N}cc}) B_0(p^2,m_c^2,m_c^2)
  + \delta_{b\H{N}} \sum_{c=1}^{\H{N}} d_{acc} (d_{\H{N}aa}
  \right. \nonumber\\
  &\left. +2
  d_{\H{N}cc}) B_0(p^2,m_c^2,m_c^2) + 4 d_{\H{N}ab} \left( d_{\H{N}aa} +
    d_{\H{N}bb} + d_{\H{N}\H{N}\H{N}} \right) B_0(p^2,m_{\H{N}}^2,m_a^2)
  \rule{0pt}{4ex} \right\},\nonumber\\
\Pi_{ab}^{(2.4)}(p^2) &= -2 g^2 \sum_{i,j=1}^N \lambda_{a_{ij}}
\lambda_{b_{ji}} \left( A_0(m_{\psi_i}^2) + A_0(m_{\psi_j}^2) +
  \left( (m_{\psi_i}+m_{\psi_j})^2 - p^2 \right)
  B_0(p^2,m_{\psi_j}^2,m_{\psi_i}^2) \rule{0pt}{2.5ex}\right).
\nonumber
\end{align}
The sums over flavor indices can be computed explicitly with the aid
of the formulas in appendix \ref{sec:appsun}, the resulting
self-energy $\Pi_{ab}(p^2)$ having the diagonal form described above.
We skip the details here, to focus only on the divergent part
$(\Pi_{ab}(p^2))_\infty$ of $\Pi_{ab}(p^2)$ and some of the finite
mass corrections.  Requiring $(\Pi_{aa}(p^2))_\infty = 0$ for the
three ranges $1\leq a \leq \H{N-1}$, $\H{N-1} < a < \H{N}$ and
$a=\H{N}$, leads to three linear relations among $\Delta \nu^2$,
$\Delta \lambda$ and $\Delta \lambda^\prime$, one of which is
identical to (\ref{eq:ct1}).  After solving those relations we find,
\begin{equation}
\begin{gathered}
  \label{eq:ctf}  
  \delta Z = - \frac{g^2}{2 \pi^2 \epsilon},
  \qquad
  \Delta \lambda = \frac{1}{8\pi^2 \epsilon} \left( (N^2+7) \lambda^2
    + 8 \frac{N^2-4}{N^2} \lambda^{\prime 2} - 4 \frac{N^2-4}{N}
    \lambda \lambda^\prime \right),\\
  \Delta \nu^2 = \frac{\nu^2}{8\pi^2 \epsilon} \left( (N^2+1) \lambda
    - 2 \frac{N^2-4}{N} \lambda^\prime \right),
    \quad
    \Delta \lambda^\prime =  \frac{1}{8\pi^2 \epsilon} \left( -4
    \frac{N^2-15}{N} \lambda^{\prime 2} + 12 \lambda \lambda^\prime
  \right). 
\end{gathered}
\end{equation}
For $N=3$ these values for $\Delta \lambda$ and $\Delta
\lambda^\prime$ are arbitrary, since the relation
$(\Pi_{aa}(p^2))_\infty = 0$ depends only on $\Delta \lambda - \Delta
\lambda^\prime/3$.  It is only this last quantity that is univocally
determined and it, as well as $\Delta \nu^2$, depends only on $\lambda
- \lambda^\prime/3$.

The finite mass corrections can be obtained explicitly from
(\ref{eq:2ptfn}).  There are two particular cases of interest. One is
the calculation of $\Pi_{aa}(p^2=0)$ with $\H{N-1} < a < \H{N}$ whose
vanishing, required by Goldstone's theorem, constitutes a non-trivial
verification of the result (\ref{eq:2ptfn}) for the self-energy.  We
omit the details of that calculation for brevity.  The other
interesting case is the computation of $\Pi_{aa}(p^2=0)$ with
$a=1,2,3$, for $N=3$, since that is the mass for $\varphi_{1,2,3}$,
which are massless at tree level.  We again omit the details and only
quote the result,
\begin{equation}
  \label{eq:wouldbe}
  \Pi_{aa}(p^2=0) = \frac{g^4 \nu^2}{\pi^2 (3\lambda -
    \lambda^\prime)} \left( \frac{8}{3} \log(2) - \frac{1}{2}\right),
  \quad
  a=1,2,3,
  \quad
  N=3.
\end{equation}
This is the common value of the renormalized masses $m^2_{1,2,3}$,
which at this order are also pole masses.  If $g=0$ those masses
vanish, as expected, since the Lagrangian without fermions has an
$SO(8)$ symmetry leading to seven Goldstone bosons.

For the fermion self-energy we have,
\begin{subequations}\label{eq:fself}
  \begin{align}
    \Pi_{ij}(p) &= \Pi_{ij}^{(V)}(p^2) \pirac + \Pi_{ij}^{(S)}(p^2),
    \quad
    1\leq i,j \leq N, \\
    \Pi_{ij}^{(V)}(p^2) &= -\delta Z_\psi \delta_{ij} + g^2
    \sum_{a=1}^{\H{N}} \sum_{l=1}^{N} (\mat{\lambda}_a)_{il}
    (\mat{\lambda}_a)_{lj} \left\{ -\frac{1}{16\pi^2 \epsilon} +
    \frac{1}{32 \pi^2} b_0(p^2,m_{\psi_l}^2,m_a^2) + \frac{1}{32 \pi^2
    p^2} 
    \right. \nonumber\\
    &\quad \times\left. \left( m_a^2 a_0(m_a^2) - m_{\psi_l}^2
        a_0(m_{\psi_l}^2) 
        + (m_{\psi_l}^2-m_a^2) b_0(p^2,m_{\psi_l}^2,m_a^2) \right)
    \rule{0pt}{3ex}\right\},\\
  \Pi_{ij}^{(S)}(p^2) &= g \Delta Z_g \left( \vect{v}_{(0)} \pdot
  \vect{\lambda} \right)_{ij} + g \left( \vect{v}_{(1)} \pdot
  \vect{\lambda} \right)_{ij} + g^2 \sum_{a=1}^{\H{N}} \sum_{l=1}^N
  (\mat{\lambda}_a)_{il} (\mat{\lambda}_a)_{lj} \left\{
    -\frac{m_{\psi_l}}{8\pi^2 \epsilon} \right.\nonumber\\
    &\quad \left.+ \frac{m_{\psi_l}}{16 \pi^2}
    b_0(p^2,m_{\psi_l}^2,m_a^2) \right\},  
  \end{align}
\end{subequations}
from whence,
\begin{equation}
  \label{eq:ctf2}
  \delta Z_\psi = -\frac{g^2}{8\pi^2 \epsilon} \frac{N^2-1}{N},
  \qquad
  \Delta Z_g = \frac{g^2}{4\pi^2 N \epsilon}.
\end{equation}
From the renormalized self-energy we can find the relation between the
renormalized and pole fermion masses. 

In order to complete the renormalization in \msb\ we have to consider
the RG evolution of the Lagrangian parameters and fields.  From the
renormalization constants given above, we find,
\begin{equation}
  \label{eq:sunbet}
\begin{gathered}
  \beta_\lambda = -\epsilon \lambda + \frac{N^2+7}{8\pi^2} \lambda^2
  + \frac{1}{\pi^2} \lambda g^2 - \frac{1}{2 \pi^2 } \frac{N^2-4}{N}
  \lambda
  \lambda^\prime + \frac{1}{\pi^2} \frac{N^2-4}{N^2} \lambda^{\prime
  2}, \\ 
  \beta_{\lambda^\prime} = -\epsilon \lambda^\prime -
  \frac{1}{2\pi^2} \frac{N^2-15}{N} \lambda^{\prime 2} +
  \frac{1}{\pi^2} \lambda^\prime g^2 + \frac{3}{2\pi^2} \lambda
  \lambda^\prime, 
  \qquad
  \beta_g = -\frac{\epsilon}{2} g + \frac{1}{16\pi^2}
  \frac{(N+1)^2}{N} g^2. 
\end{gathered}
\end{equation}
At $d=4$, $\beta_g > 0$ and, for $N \geq 3$ and $\lambda^\prime/\lambda
< 1/2$, 
\begin{equation*}
  \label{eq:posi}
  \frac{N^2+7}{8\pi^2} \lambda - \frac{1}{2 \pi^2 } \frac{N^2-4}{N}
  \lambda^\prime > 0 \quad \Rightarrow \quad \beta_\lambda > 0,
\end{equation*}
so there are no non-trivial fixed points for these coupling constants
at this order.  There is an invariant manifold, $g=0$,
$\lambda^\prime=0$, $\lambda>0$, corresponding to the $SO(\H{N})$
invariant $\lambda (\vect{\phi})^4$ theory without fermions. For
$N=3$ only $\lambda - \lambda^\prime/3$ is meaningful and from
(\ref{eq:sunbet}) we get,
\begin{equation*}
  \beta_{\lambda} - \frac{1}{3} \beta_{\lambda^\prime} = -\epsilon
  \left( \lambda - \frac{\lambda^\prime}{3} \right) + \frac{2}{\pi^2}
  \left( \lambda - \frac{\lambda^\prime}{3} \right)^2 +
  \frac{1}{\pi^2} g^2  \left( \lambda - \frac{\lambda^\prime}{3} \right),
\end{equation*}
which is also positive at $d=4$ if $\lambda^\prime/\lambda < 3$.  The
anomalous dimensions for the mass parameter $\nu^2$ and the fields are
then as follows,
\begin{equation}
  \label{eq:ano}
  \gamma_{\nu^2} = \frac{1}{2\pi^2} \left( \frac{N^2+1}{4} \lambda -
    \frac{N^2-4}{2N} \lambda^\prime + g^2 \right),
  \qquad
  \gamma_{\varphi_a} = -\frac{g^2}{4\pi^2},
  \qquad
  \gamma_\psi = -\frac{g^2}{16 \pi^2} \frac{N^2-1}{N}.
\end{equation}
The contribution to $\gamma_{\nu^2}>0$ from boson loops is positive
for $N=2$, and also for all $N>2$ if $\lambda^\prime/\lambda < 1/2$.

\subsection{Explicit symmetry breaking and non trivial mixing}
\label{sec:sunext}

A simple extension of the model (\ref{eq:sunlag}) is obtained by
substituting $-\nu^2$ in \Ll\ by a symmetric $\H{N} \times \H{N}$
mass matrix $\mat{M^2}$.  If we denote by $\mat{P}$ the $SO(\H{N})$
matrix diagonalizing $\mat{M^2}$, the new Lagrangian in the mass basis
is
\begin{equation}
  \label{eq:sunlagmat}
    \Ll = -\frac{1}{2} \vect{\phi} \pdot (\Box + \mat{M^{\prime 2}})
  \vect{\phi} - V_4(\mat{P}\vect{\phi},N) + \overline{\mat{\psi}} i
  \dirac \mat{\psi} - g \overline{\mat{\psi}} \vect{\lambda}\pdot
  \mat{P} \vect{\phi} \mat{\psi},
\end{equation}
where $\mat{M^{\prime 2}} = \diag (\nu_1^2,\dots,\nu_{\H{N}}^2)$ is
the diagonalized mass matrix.  \mat{P}\ appears now as a coupling
matrix in the Yukawa term, and also explicitly in the
non-flavor-isotropic piece of the potential for $N>3$.  If \mat{P} is
in the $SU(N)$ subgroup of $SO(\H{N})$, i.e., it is of the same form
as \mat{R}\ in (\ref{eq:suntrans}), it can be removed from the Yukawa
term by means of an $SU(N)$ transformation of the fermion fields and,
furthermore, $V_4(\mat{P}\vect{\phi},N) = V_4(\vect{\phi},N)$, so
\mat{P}\ drops from \Ll.  That is always the case if $N=2$.  In
general, for $N \geq 3$, \mat{M^2}\ needs not be diagonalizable by an
orthogonal matrix in the $SU(N)$ subgroup, and \mat{P}\ is
unremovable.  In the case $N=3$, $V_4$ is $SO(\H{N})$ invariant and,
therefore, \mat{P}\ appears only in the Yukawa term.  If $N>3$, an
$SU(N)$ transformation of fermion fields leads to a coupling matrix in
the Yukawa term that is different, though $SU(N)$ equivalent, to the
matrix \mat{P}\ in $V_4$.

The Lagrangian (\ref{eq:sunlagmat}) is approximately invariant under
the $SU(N)$ transformations (\ref{eq:suntrans}) with
$\mat{R}(\vect{\theta})$ substituted by $\mat{P}^t
\mat{R}(\vect{\theta}) \mat{P}$, only the \vect{\phi}\ mass term being
non-invariant.  Due to this softly broken $SU(N)$ symmetry the matrix
\mat{P}\ is not renormalized in \msb\ scheme.  In other schemes
\mat{P}\ may receive finite renormalization, which can be parametrized
as $\mat{P}_\naught = \mat{Q}^t \mat{P} \mat{Q}$, with $\mat{Q}$ an
orthogonal renormalization matrix (see \cite{bou} for more details
on this type of parametrization).  The treatment in this case is
completely analogous to those of the previous sections.

\section{Final remarks}
\label{sec:finale}

In section \ref{sec:1model} we consider in detail a simple model with
\ns\ scalar and \nf\ fermion fields.  
The renormalization of the model in \os\ scheme is carried out in
section \ref{sec:oss}, following the approach of \cite{bou}.  We find
the counterterms and renormalized propagators explicitly, with the
exception of contributions from diagrams with one quartic vertex,
which are given only for $\ns = 2$ in appendix \ref{sec:apploop}.
Counterterms to Yukawa couplings are also given in explicit form in
(\ref{eq:gthr}), separately for their magnitudes and unit vectors.
Whereas this can be done in closed form for row coupling matrices, it
is not possible in the case of square coupling matrices.  The
counterterms $\mat{\delta Z_H}$ and $\mat{\delta V}$ to the Hermitian
coupling matrix \mat{H}\ are only given implicitly in equation
(\ref{eq:heq}).  However, (\ref{eq:heq}) is enough to renormalize the
vertex and, in \ms, to derive RG equations for \mat{H}.  Scalar fields
in this model are unstable, so their propagators may be computed in
pole-mass scheme instead of \os. We do not expect significant changes
in the results at one loop (see, e.g., \cite{pil}).

The renormalization in \msb\ is considered in detail in section
\ref{sec:mss}.  Since the symmetry breaking effects of lower-dimension
operators do not affect wave-function and dimensionless couplings
counterterms in this scheme, those take geometric forms as in
(\ref{eq:msctr}) and (\ref{eq:msyct2}).  All counterterms and Green
functions can be obtained explicitly, with the exception of \mat{H}\ 
counterterms, like in the previous scheme.  The RG equations for the
Yukawa coupling vectors are analized in detail in \ref{sec:rgflow}.
The fixed point structure of the flow is as expected from the
symmetries of the model.  We give also a stability analysis of those
fixed points, and of the invariant manifolds.  One conclusion of that
analysis is the RG invariance of the dimension of the ``Yukawa
subspace'' of flavor space.  The results of \ref{sec:rgflow} are
independent of the masses of fermions and of other
super-renormalizable couplings in the Lagrangian, some of which we set
to zero for simplicity.  Furthermore, at one-loop level they are also
independent of the detailed form of the quartic interactions $V_4$,
and of the term $\mat{\phi}\pdot \mat{K} \mat{\phi} \chi^2$ discussed
in section \ref{sec:qlag}.

There is a close analogy between the renormalization of Yukawa versors
on one hand, and that of unitary and orthogonal matrices on the other.
In both cases the couplings take values on a compact manifold which is
covered once by the transitive action of a symmetry group, and this
property determines the form of counterterms, as in (\ref{eq:yuk})
where $\mat{\hat{a}}_{k_\naught} = \mat{Q}_{a_k} \mat{\hat{a}}_{k}$
with $\mat{Q}_{a_k}$ $\in$ $SO(\ns)$.  For this reason we expect that
the renormalization of Yukawa versors and its associated RG flow
should be a toy model of the more complicated situations in theories
with several unitary or orthogonal coupling matrices.

In section \ref{sec:2model} we consider a model with spontaneously
broken $SU(N)$ symmetry.  We choose a \textsc{vev} which is a local
minimum of the potential for all $N$, with $SU(N-1) \times U(1)$ as
stationary subgroup.  We consider the one-loop two-point function,
obtaining the masses of pseudo-Goldstone bosons in the case $N=3$, and
the \msb\ one-loop counterterms and RG equations for all $N$.  We
notice that in \msb\ the analysis of two-point functions is enough to
determine the counterterms to the coupling constants in the quartic
terms in potential, whereas in \os\ the opposite holds, namely,
renormalizing the cubic and quartic terms in the potential is
necessary in order to obtain the mass and field-strength counterterms
\cite{peskin}.

Mixing of fields is trivial in this model if the symmetry is not
explicitly broken, for instance by a mass matrix as discussed in
section \ref{sec:sunext}.  The local minima of the potential strongly
depend on the form of the mass matrix in that case.  Such explicit
breaking of the symmetry leads to the appearance in $\Ll$ of an
orthogonal coupling matrix both in $V_4$ and the Yukawa term.  It is
not possible to give closed expressions for two-point functions with a
generic mass matrix for all $N$, as done in the case without explicit
symmetry breaking.  In the \msb scheme counterterms to dimensionless
couplings are independent of the mass matrix so, in particular, the
coupling matrix $\mat{P}$ does not receive renormalization.  In a
scheme other than \msb finite counterterms to $\mat{P}$ are obtained
along the lines of the treatment of the previous sections and
\cite{bou}.

\appendix
\numberwithin{equation}{section}

\section{The potential $V_4(\mat{\phi})$}
\label{sec:appv4}

In section \ref{sec:1model} the quartic self-interactions are denoted
$V_4(\mat{\phi})$.  In this appendix we give explicit expressions for
one-loop contributions to 2- and 3- point functions from diagrams
containing one insertion of $V_4$, for the case $\ns=2$.  Higher
values of \ns\ give similar, but algebraically more involved results.
In section \ref{sec:oss} the contribution to the unrenormalized
\mat{\phi} 2-point function from those diagrams is denoted
$\mat{\Omega}^{(3)}$,
\begin{align*}
    \Omega^{(3)}_{ab} & = \frac{g}{3} \delta_{ab} \left( A_0(m_a^2)
      + \frac{1}{2} \sum_{c=1}^{2} A_0(m_c^2) \right) 
    + \frac{1}{12} \delta_{ab} \sum_{c=1}^{2} \left( \rule{0pt}{2ex}
      g_\sss{11} \mat{\hat{a}}_{1_c}^2 + g_\sss{22}
      \mat{\hat{a}}_{2_c}^2 \right)
    A_0(m_c^2) \\
    & + \frac{1}{6} \left( \rule{0pt}{2ex} g_\sss{11}
      \mat{\hat{a}}_{1_a} \mat{\hat{a}}_{1_b} + g_\sss{22}
      \mat{\hat{a}}_{2_a} \mat{\hat{a}}_{2_b} + 3 g_\sss{12}
      \left(\mat{\hat{a}}_{1_a} \mat{\hat{a}}_{2_b} +
        \mat{\hat{a}}_{1_b} \mat{\hat{a}}_{2_a} \right) \right) \left(
      A_0(m_a^2) + A_0(m_b^2) + \frac{1}{2} \sum_{c=1}^2
      A_0 (m_c^2) \right) \\
    & + \frac{g_\sss{12}}{2} \delta_{ab} \mat{\hat{a}}_1 \pdot
    \mat{\hat{a}}_2 A_0(m_a^2)
    + \frac{g_\sss{1111}}{2} \mat{\hat{a}}_{1_a} \mat{\hat{a}}_{1_b}
    \sum_{c=1}^2 \mat{\hat{a}}_{1_c}^2 A_0 (m_c^2) +
    \frac{g_\sss{2222}}{2} \mat{\hat{a}}_{2_a} \mat{\hat{a}}_{2_b}
    \sum_{c=1}^2 \mat{\hat{a}}_{2_c}^2 A_0 (m_c^2)  \\
    & + \frac{g_\sss{1122}}{12} \sum_{c=1}^2 R_{ccab} A_0(m_c^2) +
    \frac{1}{8} \sum_{c=1}^2 \left(g_\sss{1112} S_{1ccab} +
      g_\sss{1222} S_{2ccab} \right) A_0(m_c^2)
\end{align*}
where we defined the rank 4 tensors,
\begin{equation*}
  \begin{split}
    \mat{R} & = 
\mat{\hat{a}}_1 \otimes \mat{\hat{a}}_1 \otimes \mat{\hat{a}}_2
\otimes \mat{\hat{a}}_2 
+
\mat{\hat{a}}_1 \otimes \mat{\hat{a}}_2 \otimes \mat{\hat{a}}_1
\otimes \mat{\hat{a}}_2 
+
\mat{\hat{a}}_1 \otimes \mat{\hat{a}}_2 \otimes \mat{\hat{a}}_2
\otimes \mat{\hat{a}}_1 
+
\mat{\hat{a}}_2 \otimes \mat{\hat{a}}_1 \otimes \mat{\hat{a}}_1
\otimes \mat{\hat{a}}_2 \\
& +
\mat{\hat{a}}_2 \otimes \mat{\hat{a}}_1 \otimes \mat{\hat{a}}_2
\otimes \mat{\hat{a}}_1 
+
\mat{\hat{a}}_2 \otimes \mat{\hat{a}}_2 \otimes \mat{\hat{a}}_1
\otimes \mat{\hat{a}}_1 \\
     \mat{S}_1 & =
\mat{\hat{a}}_1 \otimes \mat{\hat{a}}_1 \otimes \mat{\hat{a}}_1
\otimes \mat{\hat{a}}_2 
+
\mat{\hat{a}}_1 \otimes \mat{\hat{a}}_1 \otimes \mat{\hat{a}}_2
\otimes \mat{\hat{a}}_1 
+
\mat{\hat{a}}_1 \otimes \mat{\hat{a}}_2 \otimes \mat{\hat{a}}_1
\otimes \mat{\hat{a}}_1 
+
\mat{\hat{a}}_2 \otimes \mat{\hat{a}}_1 \otimes \mat{\hat{a}}_1
\otimes \mat{\hat{a}}_1 \\
    \mat{S}_2 & = \mat{S}_1 |_{\mat{\hat{a}}_1
      \leftrightarrow \mat{\hat{a}}_2}
  \end{split}
\end{equation*}
Also in section \ref{sec:oss} the contribution to the unrenormalized 
$\mat{\phi}$--$\chi$ three-point function from diagrams with one
insertion of $V_4$ is denoted $\Gamma^{(2)}_{ab} (p_1,p_2)$.  Its
explicit form in the case $\nf =2$ is the following,
\begin{align*}
&  \Gamma^{(2)}_{ab} (p_1,p_2)  = g \mathcal{V}_{ab}^{(0)} (p_1,p_2)
  + \sum_{i\leq j=1}^{\nf} g_{ij} \mathcal{V}_{ab}^{(ij)} +
  \sum_{i\leq j\leq k\leq l=1}^{\nf} g_{ijkl}
  \mathcal{V}_{ab}^{(ijkl)}, \\
& \mathcal{V}_{ab}^{(0)} (p_1,p_2)  = -\frac{1}{3} \mu^{\frac{\epsilon}{2}}
  H_{ab} B_0((p_1+p_2)^2,m_a^2,m_b^2) - \frac{\mu^{\frac{\epsilon}{2}}}{6}
  \delta_{ab} \sum_{c=1}^{\ns} H_{cc} B_0((p_1+p_2)^2,m_c^2,m_c^2),\\
&\begin{aligned}
  \mathcal{V}_{ab}^{(11)} (p_1,p_2) & = - \frac{\mu^{\frac{\epsilon}{2}}}{6}
  \sum_{c=1}^{\ns} \mat{\hat{a}}_{1_a} \mat{\hat{a}}_{1_c} H_{bc}
  B_0((p_1+p_2)^2,m_b^2,m_c^2) \\
  & -  \frac{\mu^{\frac{\epsilon}{2}}}{6}
  \sum_{c=1}^{\ns} \mat{\hat{a}}_{1_b} \mat{\hat{a}}_{1_c} H_{ac}
  B_0((p_1+p_2)^2,m_a^2,m_c^2) 
   -\frac{\mu^{\frac{\epsilon}{2}}}{12}
  \sum_{c=1}^{\ns} \mat{\hat{a}}_{1_a} \mat{\hat{a}}_{1_b} H_{cc}
  B_0((p_1+p_2)^2,m_c^2,m_c^2)\\
  &-\frac{\mu^{\frac{\epsilon}{2}}}{12} \delta_{ab}
  \sum_{c,d=1}^{\ns} \mat{\hat{a}}_{1_c} \mat{\hat{a}}_{1_d} H_{cd}
  B_0((p_1+p_2)^2,m_c^2,m_d^2),
 \end{aligned}  \\
& \mathcal{V}_{ab}^{(22)} (p_1,p_2) = \left. \mathcal{V}_{ab}^{(11)}
  (p_1,p_2) \right|_{\mat{\hat{a}}_1 \rightarrow \mat{\hat{a}}_2},\\
&\begin{gathered}
  \mathcal{V}_{ab}^{(12)} (p_1,p_2) = -g_{12} \left\{ \frac{1}{24}
  \mat{\hat{a}}_{1_a} \mat{\hat{a}}_{2_b} \sum_{c=1}^{\ns} H_{cc}
  B_0((p_1+p_2)^2,m_c^2,m_c^2) +
  \frac{1}{12}
  \sum_{c=1}^{\ns} H_{cb} \left( \mat{\hat{a}}_{1_a}
  \mat{\hat{a}}_{2_c} + \mat{\hat{a}}_{1_c} \mat{\hat{a}}_{2_a}  \right)
   \right. \\
  \left. \times B_0((p_1+p_2)^2,m_c^2,m_b^2) + 
  \frac{1}{24} \delta_{ab}
  \sum_{c,d=1}^{\ns} H_{cd} \mat{\hat{a}}_{1_c}
  \mat{\hat{a}}_{2_d} B_0((p_1+p_2)^2,m_c^2,m_d^2) \right\} +
(a\leftrightarrow b),
 \end{gathered}\\
& \mathcal{V}_{ab}^{(1111)} (p_1,p_2)  = - \frac{\mu^{\frac{\epsilon}{2}}}{2}
  \mat{\hat{a}}_{1_a} \mat{\hat{a}}_{1_b} 
  \sum_{c,d=1}^{\ns} \mat{\hat{a}}_{1_c} \mat{\hat{a}}_{1_d} H_{cd}
  B_0((p_1+p_2)^2,m_c^2,m_d^2),\\
& \mathcal{V}_{ab}^{(2222)} (p_1,p_2)  = \left. \mathcal{V}_{ab}^{(1111)}
  (p_1,p_2) \right|_{\mat{\hat{a}}_1 \rightarrow \mat{\hat{a}}_2},\\
& \mathcal{V}_{ab}^{(1122)} (p_1,p_2)  =   -\frac{\mu^{\frac{\epsilon}{2}}}{12}
  \sum_{c,d=1}^{\ns} R_{abcd} H_{cd} B_0((p_1+p_2)^2,m_c^2,m_d^2),
  \quad \text{with $\mat{R}$ defined above},\\
& \mathcal{V}_{ab}^{(1112)} (p_1,p_2)  =  -\frac{\mu^{\frac{\epsilon}{2}}}{8}
  \sum_{c,d=1}^{\ns} S_{1 abcd} H_{cd} B_0((p_1+p_2)^2,m_c^2,m_d^2),
  \quad \text{with $\mat{S}_1$ defined above},\\
& \mathcal{V}_{ab}^{(1222)} (p_1,p_2)  = \left. \mathcal{V}_{ab}^{(1112)}
  (p_1,p_2) \right|_{\mat{\hat{a}}_1 \rightarrow \mat{\hat{a}}_2}.
\end{align*}
Gathering the $\epsilon$-poles, we obtain the divergent piece of the
\mat{H} counterterms,
\begin{equation} \label{eq:dhapp}
  \begin{gathered}
  (8\pi^2 \epsilon) \mat{\Delta H}  = \frac{g}{3} \left( \mat{H} +
    \frac{1}{2} \Tr{\mat{H}} \right)
   + \left( \frac{g_{1111}}{2} \left( \mat{\hat{a}}_1 \pdot
   \mat{H}\mat{\hat{a}}_1 \mat{\hat{a}}_1 \otimes \mat{\hat{a}}_1
   \right) +   \frac{g_{2222}}{2} \left( \mat{\hat{a}}_1 \rightarrow
   \mat{\hat{a}}_2 \right) \right) \\
  + \frac{g_{11}}{6} \left( \mat{\hat{a}}_1 \otimes
   \mat{H}\mat{\hat{a}}_1 +
   \mat{H}\mat{\hat{a}}_1 \otimes \mat{\hat{a}}_1 +
   \frac{1}{2}\Tr{\mat{H}}\mat{\hat{a}}_1 \otimes \mat{\hat{a}}_1 +
   \frac{1}{2}\mat{\hat{a}}_1 \pdot \mat{H}\mat{\hat{a}}_1\right) +
 \frac{g_{22}}{6} \left( \mat{\hat{a}}_1 \rightarrow \mat{\hat{a}}_2
   \right)\\
    + \frac{g_{12}}{12} \left( \mat{\hat{a}}_1 \otimes
   \mat{H}\mat{\hat{a}}_2 +
   \mat{H}\mat{\hat{a}}_2 \otimes \mat{\hat{a}}_1 +
   \frac{1}{2}\Tr{\mat{H}}\mat{\hat{a}}_1 \otimes \mat{\hat{a}}_2 +
   \frac{1}{2}\mat{\hat{a}}_1 \pdot \mat{H}\mat{\hat{a}}_2\right) +
 \frac{g_{12}}{12} \left( \mat{\hat{a}}_1 \leftrightarrow \mat{\hat{a}}_2
   \right)\\
  + \frac{g_{1112}}{8} \left( (\mat{\hat{a}}_1 \pdot
   \mat{H}\mat{\hat{a}}_2 + \mat{\hat{a}}_2 \pdot
   \mat{H}\mat{\hat{a}}_1) \mat{\hat{a}}_1 \otimes
   \mat{\hat{a}}_1 + \mat{\hat{a}}_1 \pdot \mat{H}\mat{\hat{a}}_1 (
   \mat{\hat{a}}_1 \otimes \mat{\hat{a}}_2 + \mat{\hat{a}}_2 \otimes
   \mat{\hat{a}}_1) \right)
 +   \frac{g_{1222}}{8} \left( \mat{\hat{a}}_1 \leftrightarrow
   \mat{\hat{a}}_2 \right) 
\end{gathered}
\end{equation}

\subsection{One-loop renormalization of $V_4(\mat{\phi})$}
\label{sec:v4ct}

In this section we briefly discuss the form of counterterms to $V_4$
in \ms\ scheme and check that those counterterms have the same form as
$V_4$ at one-loop level.  The explicit computation of counterterms to
$V_4$ in other schemes, such as \os, can be somewhat involved due to
their finite parts violating \zy\ symmetry.  The form of the potential
$V_4$ in the classical theory is (\ref{eq:pot}).  The potential and
its counterterms in terms of renormalized fields is given in
(\ref{eq:cpot}), which can be rewritten as
\begin{equation}
  \label{eq:ccpot}
    V_4 \left(\mat{Z}^{1/2} \mat{\phi}, \{\mat{U}_{k}
    \mat{\hat{a}}_{k}\}, \mu^\epsilon (g + \delta g),\{\mu^\epsilon
    (g_{ij} + \delta g_{ij})\}, \{\mu^\epsilon (g_{ijkl} + \delta
    g_{ijkl})\}\right),
\end{equation}
with $\mat{U}_k \equiv \mat{Q}^t \mat{Q}_{a_k}$ and, in \ms,
$\mat{Q} = \mat{1}$.  At one-loop level, (\ref{eq:ccpot}) has
the form,
  \begin{multline}  \label{eq:cccpot}
    V_4 \left(\mat{Z}^{1/2} \mat{\phi}, \{\mat{U}_{k}
      \mat{\hat{a}}_{k}\}, \mu^\epsilon g,\{\mu^\epsilon
      g_{ij}\}, \{\mu^\epsilon g_{ijkl}\} \right) \\
    + V_4 \left(\mat{\phi}, \{\mat{\hat{a}}_{k}\}, \mu^\epsilon \delta
      g,\{\mu^\epsilon \delta g_{ij})\}, \{\mu^\epsilon \delta
      g_{ijkl})\}\right) + \text{higher orders}
  \end{multline}
The second term has the same form as $V_4$.  We want to see that
this is also the case for the first term, whose explicit expression
is,
\begin{multline}
  \label{eq:ccccpot}
  \frac{g}{4!} \left( \mat{\phi} \pdot \mat{Z} \mat{\phi} \right)^2 +
  \frac{1}{4!} \sum_{i\leq j=1}^{\nf} g_{ij} \left( \mat{\hat{a}}_i
    \pdot \mat{U}^t_i \mat{Z}^{1/2} \mat{\phi} \right) \left(
    \mat{\hat{a}}_j \pdot \mat{U}^t_j \mat{Z}^{1/2} \mat{\phi} \right)
  \left( \mat{\phi} \pdot \mat{\delta Z} \mat{\phi} \right) \\
  + \frac{1}{4!} \sum_{i\leq j\leq k\leq l=1}^{\nf} g_{ijkl} \left(
    \mat{\hat{a}}_i \pdot \mat{U}^t_i \mat{Z}^{1/2} \mat{\phi} \right)
  \left( \mat{\hat{a}}_j \pdot \mat{U}^t_j \mat{Z}^{1/2} \mat{\phi}
  \right) \left( \mat{\hat{a}}_k \pdot \mat{U}^t_k \mat{Z}^{1/2}
    \mat{\phi} \right) \left( \mat{\hat{a}}_l \pdot \mat{U}^t_l
    \mat{Z}^{1/2} \mat{\phi} \right).
\end{multline}
The one-loop value for \mat{\delta Z} is given in (\ref{eq:msctr}) and
for $\mat{\delta U_j} = \mat{\delta Q}_{a_j}$ in
(\ref{eq:msyct2}).  Substituting those expressions in
(\ref{eq:ccccpot}) we get
\begin{multline}\label{eq:v4tc}
  V_4 \left(\mat{\phi}, \{\mat{\hat{a}}_{k}\}, \mu^\epsilon
    g,\{\mu^\epsilon g_{ij}\}, \{\mu^\epsilon g_{ijkl}\}\right) -
  \frac{g}{4! 2\pi^2 \epsilon} \sum_{k=1}^{\nf} a_k^2
  (\mat{\hat{a}}_k \pdot \mat{\phi})^2 \mat{\phi}^2 \\
  - \frac{1}{4! 8\pi^2 \epsilon} \sum_{i,j=1}^{\nf} g_{ij} \left(
    \sum_{m=1}^{\nf} a_m^2 \left(x_{mi}^2 + x_{mj}^2\right) \right)
  \mat{\hat{a}}_i \pdot \mat{\phi}\, \mat{\hat{a}}_j \pdot
  \mat{\phi}\,\mat{\phi}^2 
  - \frac{1}{4! 4\pi^2 \epsilon} \sum_{i,j,k=1}^{\nf} g_{ij} a_k^2
  \mat{\hat{a}}_i \pdot \mat{\phi}\, \mat{\hat{a}}_j \pdot \mat{\phi}
  \left( \mat{\hat{a}}_k \pdot \mat{\phi} \right)^2 \\
  - \frac{1}{4! 8\pi^2 \epsilon} \sum_{i,j,k,l=1}^{\nf} g_{ijkl}
  \left( \sum_{m=1}^{\nf} a_m^2 \left(x_{mi}^2 + x_{mj}^2 + x_{mk}^2 +
      x_{ml}^2\right) \right) \mat{\hat{a}}_i \pdot \mat{\phi}\,
  \mat{\hat{a}}_j \pdot \mat{\phi}\, \mat{\hat{a}}_k \pdot
  \mat{\phi}\, \mat{\hat{a}}_l \pdot \mat{\phi}
  + \mathrm{h.o.},
\end{multline}
where h.o.\ refers to terms of higher order.  (\ref{eq:v4tc}) clearly
has the same form as (\ref{eq:pot}).

\section{Loop integrals}
\label{sec:apploop}

In this appendix we give a list of loop integrals used in the
foregoing.  More complete calculations can be found, e.g., in
\cite{den,vel,pve,ste}.  Divergent integrals are separated in a
dimensional regularization pole term and a finite remainder.  $\mub =
\mu\sqrt{4\pi e^{-\gamma_E}}$.
\begin{align*}
&
  A_0(m^2) = \frac{i\mu^\epsilon}{(2\pi)^d} \int d^d\ell
  \frac{1}{\ell^2-m^2+i\varepsilon} = -\frac{m^2}{8\pi^2\epsilon} +
  \frac{m^2}{16\pi^2} a_0(m^2),
  \quad
  a_0(m^2) = \log\left(\frac{m^2}{\mub^2}\right) - 1\\
& A_1^\mu(m^2) = \frac{i\mu^\epsilon}{(2\pi)^d} \int d^d\ell
  \frac{\ell^\mu}{\ell^2-m^2+i\varepsilon} = 0\\
&
  A_2^{\mu\nu}(m^2) = \frac{i\mu^\epsilon}{(2\pi)^d} \int d^d\ell
  \frac{\ell^\mu\ell^\nu}{\ell^2-m^2+i\varepsilon} =
  -\frac{m^4}{32\pi^2\epsilon} g^{\mu\nu} + \frac{m^4}{64\pi^2}
  g^{\mu\nu} a_2(m^2),
  \quad
  a_2(m^2) = a_0(m^2) - \frac{1}{2}
  \\
& A_2(m^2) = \frac{i\mu^\epsilon}{(2\pi)^d} \int d^d\ell
  \frac{\ell^2}{\ell^2-m^2+i\varepsilon} = m^2 A_0(m^2)\\
& 
\begin{aligned}
B_0(p^\mu,m_1^2,m_2^2) & = \frac{i\mu^\epsilon}{(2\pi)^d} \int d^d\ell
  \frac{1}{\left(\ell^2-m_1^2+i\varepsilon\right)
    \left((\ell+p)^2-m_2^2+i \varepsilon\right)}
   = -\frac{1}{8\pi^2\epsilon} + \frac{1}{16\pi^2}
   b_0(p^2,m_1^2,m_2^2) \\
  b_0(p^2,m_1^2,m_2^2) &= \int_0^1 dx \log\left((1-x)
  \frac{m_1^2}{\mub^2} + x \frac{m_2^2}{\mub^2} - x(1-x)
  \frac{p^2}{\mub^2} -i\varepsilon\right)  
\end{aligned}\\
&
\begin{aligned}
  B_1^\mu(p^\mu,m_1^2,m_2^2) &= \frac{i\mu^\epsilon}{(2\pi)^d} \int d^d\ell
  \frac{\ell^\mu}{\left(\ell^2-m_1^2+i\varepsilon\right)
    \left((\ell+p)^2-m_2^2+i \varepsilon\right)}
  = \frac{p^\mu}{16\pi^2\epsilon} - \frac{p^\mu}{16\pi^2}
  b_1(p^2,m_1^2,m_2^2) \\
  b_1(p^2,m_1^2,m_2^2) &= \int_0^1 dx x \log\left((1-x)
  \frac{m_1^2}{\mub^2} + x \frac{m_2^2}{\mub^2} - x(1-x)
  \frac{p^2}{\mub^2} -i\varepsilon\right)
\end{aligned}\\
\intertext{which can also be written,}
&\begin{aligned}
  B_1^\mu(p^\mu,m_1^2,m_2^2) &= p^\mu B_1(p^2,m_1^2,m_2^2)\\
  p^2 B_1(p^2,m_1^2,m_2^2) &= \frac{1}{2}\left(A_0(m_1^2) - A_0(m_2^2)
    - (p^2+m_1^2-m_2^2) B_0(p^2,m_1^2,m_2^2)\right)
\end{aligned}\\
&  C_0(p_1,p_2,m_1^2,m_2^2,m_3^2)=
   \frac{i\mu^\epsilon}{(2\pi)^d} 
   \int \negthickspace d^d\ell
  \frac{1}{(\ell^2-m_1^2+i\varepsilon)
  ((\ell+p_2)^2-m_2^2+i \varepsilon)
  ((\ell-p_1+p_2)^2-m_3^2+i \varepsilon)}\\
\end{align*}

\section{Counterterms to a regular coupling matrix}
\label{sec:appmat}

The counterterms \mat{\delta Z_H} and \mat{\delta V} to the coupling
matrix \mat{H}, as defined in (\ref{eq:counter}), can be found by
solving (\ref{eq:heq}) along the lines of appendix C of \cite{bou},
where the more general case of complex normal matrices \mat{H} is
considered.  In this appendix we point out that the solution can be
simplified if \mat{H} is regular.  Our expansion parameters are the
(dimensionful) eigenvalues of \mat{H}.  In the basis in which \mat{H}
is diagonal, $\mat{H} = \diag (h_1,\dots, h_{\ns})$, we have $h_j
\rightarrow 0$, $h_i/h_j = \mathcal{O}(1)$.  We define,
\[
\varepsilon = \sqrt{|h_1|^2 + \dots +|h_{\ns}|^2} = \sqrt{\tr\left(
    \mat{H}^\dagger \mat{H}\right)} = ||\mat{H}||,
\]
as our perturbation parameter.  We want to find \mat{\delta Z_H} from
the equation,
\begin{equation}
  \label{eq:heqa}
  \mat{\delta Z_H} \mat{H} + \left[ \mat{\delta V}, \mat{H} \right] =
  \mat{X}, \quad
  \left[ \mat{\delta Z_H}, \mat{H} \right] = 0.
\end{equation}
Here, \mat{X}\ stands for a known matrix.  In the case considered in
this paper, \mat{X}\ is given by the r.h.s.\ of (\ref{eq:heq}) and it
is real symmetric.  Since we assume \mat{H}\ to be regular,
\mat{\delta Z_H} must be a linear combination of $\mat{H}^k$ with
$k=0,\dots,\ns-1$.  At one-loop level, $\mat{\delta Z_H} =
\mathcal{O}(\varepsilon^2)$ so we write,
\[
\mat{\delta Z_H} = \delta z_0 \mat{1} + \delta z_1 \mat{H} + \delta
z_2 \mat{H}^2 + \mathcal{O}(\varepsilon ^3).
\] 
In order to project (\ref{eq:heqa}), we orthogonalize the basis
$\{\mat{1},\mat{H},\mat{H}^2\}$ of matrices commuting with \mat{H} and
being $\mathcal{O}(\varepsilon^2)$.  The orthogonal set has the form
$\mat{N_0} = \mat{1}$, $\mat{N_1} = \mat{H} - 1/\ns \tr(\mat{H})
\mat{1}$, $\mat{N_2} = \mat{H}^2 - \nu_1 \mat{H} + \nu_0 \mat{1}$,
with $\nu_{1,2}$ appropriately determined.  When \mat{H} is normal,
$\tr \left( \left[ \mat{\delta V}, \mat{H} \right] \mat{H}^n
  \mat{H}^{\dagger m}\right) = 0$ for $n,m \geq 0$.  Thus, the
projections of $\left[ \mat{\delta V}, \mat{H}\right]$ over
$\mat{N_{0,1,2}}$ vanish, and we can obtain \mat{\delta Z_H} by
orthogonal projection from (\ref{eq:heqa}).

\section{Decomposition of a vector function}
\label{sec:appdec}

Assume we are given a function $\mat{f}: \mathbb{R}^N \rightarrow
\mathbb{R}^{N}$ of the form $\mat{f}(\mat{x}) = \mat{A}\mat{x}$, where
$\mat{A} \in \mathbb{R}^{N\times N}$ can be \mat{x} dependent.  We can
always find a scalar function $\lambda = \lambda (\mat{x}) \geq 0$ and
an orthogonal matrix $\mat{Q} = \mat{Q}(\mat{x}) \in SO(N)$ such that
$\mat{f}(\mat{x}) = \lambda \mat{Q x}$.
If the matrix \mat{A} is close to the identity, $\mat{A} = \mat{1} +
\mat{\delta A}$, to lowest non-trivial order in \mat{\delta A} we
have,
\[
\lambda = 1 + \delta \lambda,
\quad
\delta \lambda = \mat{\hat{x}} \pdot \mat{\delta A \hat{x}},
\quad
\mat{Q} = \exp (-\mat{\delta Q}),
\quad
\mat{\delta Q} = (\mat{\delta A \hat{x}}) \wedge \mat{\hat{x}},
\]
where $\mat{\hat{x}} = \mat{x}/||\mat{x}||$, and $\mat{u} \wedge
\mat{v} = \mat{u} \otimes \mat{v} - \mat{v} \otimes \mat{u}$.  It
is clear that $\delta \lambda \mat{x} + \mat{\delta Q} \mat{x}
= \mat{\delta A} \mat{x}$.

\section{RG fixed points }
\label{sec:appbeta}

The evolution of Yukawa versors under the RG flow is discussed in
section \ref{sec:rgflow}.  In this appendix we consider fixed points
of the third type, and show that they are saddle points of the flow.
The beta function for the Gram matrix \mat{x} is given in
(\ref{eq:xbet}).  From there, we obtain,
\begin{align}
    8\pi^2 \frac{\partial \beta_{x_{ij}}}{\partial x_{kl}} & =
    2 a_k^2 \left( \delta_{il} (x_{kj} - x_{ki} x_{ij}) + \delta_{jl}
      (x_{ki} - x_{kj} x_{ij}) \right) 
  + 2 a_l^2 \left( \delta_{ik} (x_{lj} - x_{li} x_{ij}) + \delta_{jk}
    (x_{li} - x_{lj} x_{ij}) \right) \nonumber\\
   & - \sum_{m=1}^{\nf} a_m^2 \left( x_{mi}^2 + x_{mj}^2 \right) \left(
    \delta_{ik} \delta_{jl} + \delta_{il} \delta_{jk}
    \right).\label{eq:dxbet} 
\end{align}
We consider \mat{\beta_x} as a function of the $\binom{\nf}{2}$
variables $(x_{ij})_{i<j}$.  Thus, $\mat{\beta_x} :
\mathbb{R}^{\binom{\nf}{2}} \rightarrow \mathbb{R}^{\binom{\nf}{2}}$,
and (\ref{eq:dxbet}) are the entries of a $\binom{\nf}{2}\times
\binom{\nf}{2}$ matrix.

At a fixed point of the third type, the Yukawa versors lie on
orthogonal directions, with at least two of them being collinear
($x_{ij} = 1$ for some $i<j$), and at least two being orthogonal
($x_{kl} = 0$ for some $k<l$).  We can always relabel
$\{\mat{\hat{a}}_k\}$ so that $\{ \mat{\hat{a}}_1, \ldots,
\mat{\hat{a}}_{n_1}\}$ are collinear, then along an orthogonal
direction $\{ \mat{\hat{a}}_{n_1+1}, \ldots,
\mat{\hat{a}}_{n_1+n_2}\}$ are collinear, and so on.  If the $\{
\mat{\hat{a}}_k \}$ lie along $r$ different orthogonal directions in
flavor space, $n_1 + \ldots + n_r = \nf$.

With that convention, the matrix \mat{\xo} has $r$ diagonal blocks of
dimension $n_i \times n_i$, $i=1,\ldots, r$.  In each of these
diagonal blocks all entries are equal to 1, and outside the diagonal
blocks all entries are 0.  Notice that we have chosen all
$x_{ij}^\naught \geq 0$, as explained in section \ref{sec:rgflow}.  It
is easy to check that such \mat{\xo} satisfies (\ref{eq:geo}).  We
consider first the case of two orthogonal directions, $r=2$.  The
generalization to $r>2$ presents no difficulty and is discussed below.

In the case $r=2$, \mat{\xo} has two diagonal blocks of sizes $n_1$,
$n_2$ with $n_1+n_2=\nf$.  Assuming $n_{1,2} > 1$ for the moment, from
(\ref{eq:dxbet}) we find that $\partial \beta_{ij}/\partial x_{kl}$
has three diagonal blocks, outside of which all entries are 0.  The
first one consists of the entries $\partial \beta_{ij}/\partial
x_{kl}$ with $ 1 \leq i < j \leq n_1$ and $ 1 \leq k < l \leq n_1$.
This submatrix of dimensions $\binom{n_1}{2} \times \binom{n_1}{2}$ is
diagonal, with diagonal entries all equal to $-2 \sum_{m=1}^{n_1}
a_m^2 < 0$.

There is another diagonal block consisting of $\partial
\beta_{ij}/\partial x_{kl}$ with $ n_1+1 \leq i < j \leq n_1 + n_2$
and $ n_1+1 \leq k < l \leq n_1 + n_2$, of size
$\binom{n_2}{2} \times \binom{n_2}{2}$ , which is
also diagonal with all diagonal entries equal to $-2
\sum_{m=n_1+1}^{n_1+n_2} a_m^2$.  If $n_2 =1$ this block is not
present.

There is, finally, another diagonal block consisting of $\partial
\beta_{ij}/\partial x_{kl}$ with $ 1 \leq i , k \leq n_1$ and $ n_1+1
\leq j , l \leq n_1 + n_2$.  This block has size $(n_1 n_2) \times
(n_1 n_2)$ (notice that $\binom{n_1}{2} + \binom{n_2}{2} + n_1 n_2 =
\binom{\nf}{2}$).  From (\ref{eq:dxbet}), with the explicit form for
\mat{\xo}\ given above, we obtain,
\begin{equation}
  \label{eq:fxpt}
  8\pi^2 \sum_{k=1}^{n_1} \sum_{l=n_1+1}^{n_1+n_2} \left(\frac{\partial
  \beta_{x_{ij}}}{\partial x_{kl}}\right)_{\mat{\xo}} =
  \sum_{m=1}^{n_1+n_2} a_m^2,
  \quad
  1 \leq i \leq n_1,
  \quad
  n_1+1 \leq j \leq n_1+n_2.
\end{equation}
Thus, the sum of the entries of any row of this submatrix is the
same for all rows, given by the r.h.s.\ of (\ref{eq:fxpt}) which is
independent of $(i,j)$.  This implies that we can write this submatrix
as, 
\begin{equation}
  \label{eq:subm}
  \mat{A} + \left( \sum_{m=1}^{n_1 + n_2} a_m^2 \right) \mat{1},
\end{equation}
where the $(n_1 n_2) \times (n_1 n_2)$ matrix \mat{A} is such that the
sum of its columns vanishes.  We conclude that \mat{A} has a null
eigenvalue and then, from (\ref{eq:subm}), the diagonal block we are
considering must have $\sum_{m=1}^{n_1 + n_2} a_m^2 > 0$ as
eigenvalue.  The other eigenvalues of this block can be positive,
negative or zero, depending on the values of the couplings $a_k$.

Since we are considering a fixed point \mat{\xo} of third type, at
least one of $n_1$, $n_2$ must be $> 1$, and we can always assume that
$n_1 > 1$.  From the foregoing, we see that independently of the
values of the coupling constants $a_k$, and of whether $n_2 =1$ or
$n_2 > 1$, the linearized beta function has at least one negative
eigenvalue $-2 \sum_{m=1}^{n_1} a_m^2$, and at least one positive
eigenvalue $\sum_{m=1}^{n_1+n_2} a_m^2$.  Thus, these fixed points are
saddle points.

The generalization of this conclusion to the case $r > 2$ is
immediate.  We have diagonal blocks of size $\binom{n_i}{2} \times
\binom{n_i}{2}$ with eigenvalues $-2 \sum_{m = n_1 + \ldots + n_{i-1}
  + 1}^{n_1 + \ldots + n_i}a_m^2$ if $n_i > 1$, and blocks of size
$(n_i n_j) \times (n_i n_j)$, $i \neq j$, with one eigenvalue equal to
$ \sum_{m=n_1+\ldots + n_{i-1}+1}^{n_1+\ldots +n_i} a_m^2 +
\sum_{m=n_1+\ldots + n_{j-1}+1}^{n_1+\ldots +n_j} a_m^2$. (By
convention we set $n_0 =1$, if it occurs in these expressions.)  In
this general case, too, the linearized beta function has negative and
positive eigenvalues at a fixed point of third type.

\section{The \mat{su(N)} algebra}
\label{sec:appsun}

In this appendix we gather our notations and conventions for the
$su(N)$ algebra, and several relations among its constants that are
used in section \ref{sec:2model}.  
We denote $\H{k}\equiv k^2-1$ for brevity.  Gell-Mann matrices for
$su(N)$ are denoted $\mat{\lambda}^{(N)}_a$, $1 \leq a \leq \H{N}$, with
the supraindex indicating the dimension usually omitted.  The diagonal
matrices are
\begin{equation*}
  \mat{\lambda}^{(N)}_{\H{m}} = \sqrt{\frac{2}{m(m-1)}} \diag (
  \underbrace{1,\dots, 1}_m, -(m-1),0,\dots,0),
  \qquad
  m=2,\dots,N.
\end{equation*}
The definition we give for $\mat{\lambda}^{(N)}_a$ is recursive.
$\mat{\lambda}^{(2)}_{1,2,3}$ are Pauli matrices.  Given the set
$\{\mat{\lambda}^{(N)}_a\}_{a=1}^{\H{N}}$, we obtain
$\mat{\lambda}^{(N+1)}_a$ with $a \leq \H{N}$ by embedding
$\mat{\lambda}^{(N)}_a$ in the upper-left corner of a null
$(N+1)\times (N+1)$ matrix.  The matrix
$\mat{\lambda}^{(N+1)}_{\H{N+1}}$ is diagonal, as defined above.
Finally, the remaining $2N$ matrices $\mat{\lambda}^{(N+1)}_a$ with
$\H{N} < a < \H{N+1}$ are,
\begin{multline*}
\mat{\lambda}^{(N+1)}_{N^2} = 
  \begin{pmatrix}
  0     & 0     &\ldots& 0     &  1      \\
  0     & 0     &\ldots& 0     &  0      \\
  \vdots& \vdots&      & \vdots&  \vdots \\
  0     & 0     &\ldots& 0     &  0      \\
  1     & 0     &\ldots& 0     &  0      \\   
  \end{pmatrix},\quad
\mat{\lambda}^{(N+1)}_{N^2+1} = 
  \begin{pmatrix}
  0     & 0     &\ldots& 0     &  -i     \\
  0     & 0     &\ldots& 0     &  0      \\
  \vdots& \vdots&      & \vdots&  \vdots \\
  0     & 0     &\ldots& 0     &  0      \\
  i     & 0     &\ldots& 0     &  0      \\   
\end{pmatrix}, \ldots\\
\ldots, \mat{\lambda}^{(N+1)}_{\H{N+1}-2} = 
  \begin{pmatrix}
  0     & 0     &\ldots& 0     &  0      \\
  \vdots& \vdots&      & \vdots&  \vdots \\
  0     & 0     &\ldots& 0     &  0      \\
  0     & 0     &\ldots& 0     &  1      \\
  0     & 0     &\ldots& 1     &  0      \\   
  \end{pmatrix},\quad
\mat{\lambda}^{(N+1)}_{\H{N+1}-1} = 
  \begin{pmatrix}
  0     & 0     &\ldots& 0     &  0      \\
  \vdots& \vdots&      & \vdots&  \vdots \\
  0     & 0     &\ldots& 0     &  0      \\
  0     & 0     &\ldots& 0     &  -i     \\
  0     & 0     &\ldots& i     &  0      \\   
\end{pmatrix}.
\end{multline*}
We have the usual relations,
\begin{equation}
  \label{eq:rel}  
\begin{gathered}
  \sum_{a=1}^{\H{N}} \mat{\lambda}_a \mat{\lambda}_a = 2 \frac{N^2-1}{N}
  \mat{1}, \quad
  \sum_{a=1}^{\H{N}} \mat{\lambda}_a \mat{\lambda}_b \mat{\lambda}_a =
  -\frac{2}{N} \mat{\lambda}_b, \quad
  \sum_{a=1}^{\H{N}} (\mat{\lambda}_a)_{ij} (\mat{\lambda}_a)_{kl} = 2
  \delta_{il} \delta_{kj} - \frac{2}{N} \delta_{ij} \delta_{kl},\\
  \Tr{\mat{\lambda}_a \mat{\lambda}_b} = 2 \delta_{ab}, \quad
  \Tr{\mat{\lambda}_a \mat{\lambda}_b \mat{\lambda}_c} = 2 (d_{abc} + i
  f_{abc}).
\end{gathered}
\end{equation}
The totally antisymmetric structure constants $f_{abc}$ and the
totally symmetric constants $d_{abc}$ are defined relative to this
basis by,
\begin{equation}
  \label{eq:sunfd}
  \mat{\lambda}_a \mat{\lambda}_b = \sum_{c=1}^{\H{N}} (d_{abc} + i
  f_{abc}) \mat{\lambda}_c + \frac{2}{N} \delta_{ab} \mat{1}.
\end{equation}
Particularly important for the evaluation of radiative corrections in
section \ref{sec:2model} are the constants $f_{abc}$ and $d_{abc}$
with one index equal to $\H{N}$.  Those are given by,
\begin{equation}
  \label{eq:fh}
  f_{\H{N}ab} = \begin{cases}
      \sqrt{\frac{N}{2(N-1)}} & \text{if $\H{N-1} < a < \H{N}$, 
      $b=a+1$,  and $\mat{\lambda}_a$  is real,} \\
      -\sqrt{\frac{N}{2(N-1)}} &  \text{if $\H{N-1} < a < \H{N}$, 
      $b=a-1$,  and $\mat{\lambda}_a$ is imaginary,} \\
      0  & \text{otherwise,}
    \end{cases}
\end{equation}
and $d_{\H{N}ab} = d_{\H{N}aa} \delta_{ab}$ with,
\begin{equation}
  \label{eq:dh}
  d_{\H{N}aa} = \begin{cases}
       \sqrt{\frac{2}{N(N-1)}} & \text{if  $1 \leq a \leq \H{N-1}$,} \\
   -\frac{N-2}{\sqrt{2N(N-1)}} & \text{if   $\H{N-1} < a < \H{N}$,}\\
   -(N-2) \sqrt{\frac{2}{N(N-1)}} & \text{if $a = \H{N}$.}
    \end{cases}
\end{equation}
With these equations we can compute the algebraic expressions
appearing in loop diagrams in section \ref{sec:2model}.  We quote here
the results for the divergent parts.  In fermion tadpoles we have,
with $m_{\psi_i}$ given in (\ref{eq:fmass}),
\begin{equation}
  \label{eq:ftad}
  \sum_{i=1}^N (\mat{\lambda}_{\H{N}})_{ii} m_{\psi_i}^3 = 4
  \frac{N^2-3N+3}{N(N-1)} (g v_{(0)})^3,
\end{equation}
and in fermion bubble diagrams,
\begin{align}
  \lefteqn{\sum_{i,j=1}^N (\mat{\lambda}_a)_{ij}
    (\mat{\lambda}_b)_{ji} (m_{\psi_i}^2 +
    m_{\psi_j}^2 + m_{\psi_i} m_{\psi_j}) =} \nonumber\\
  &= (g v_{(0)})^2 \delta_{ab} 2 \left( \frac{4}{N} + (d_{\H{N}aa})^2 +
    2 d_{\H{N}aa} d_{\H{N}\H{N}\H{N}} - (f_{\H{N}a(a+1)})^2 -
    (f_{\H{N}a(a-1)})^2 + \frac{2}{N} \delta_{a\H{N}} \right)\\
  &= (g v_{(0)})^2 \delta_{ab} \frac{1}{N(N-1)} \times
    \begin{cases}
      12 & \text{if $1 \leq a \leq \H{N-1}$} \\
      4 (N^2-3N+3)  & \text{if  $\H{N-1} < a < \H{N}$} \\
      12 (N^2-3N+3) & \text{if  $a=\H{N}$}
    \end{cases}. \nonumber
\end{align}
In boson tadpole diagrams we find expressions of the form,
\begin{equation}
  \label{eq:btad}
  \sum_{c=1}^{\H{N}} d_{\H{N}cc} (d_{\H{N}\H{N}\H{N}} + 2 d_{\H{N}cc})
  m_c^2 = -2 \frac{(N-2) (N-4)}{N-1} m_1^2 + 6 \frac{(N-2)^2}{N(N-1)}
  m_{\H{N}}^2, 
\end{equation}
with $m_a^2$ from (\ref{eq:smass}).  In boson bubble diagrams we find
expressions of the form,
\begin{flalign}
&  \sum_{c,d=1}^{\H{N}} d_{\H{N}aa} d_{\H{N}bb} d_{acd} d_{bcd} =
  \frac{N^2-4}{N} (d_{\H{N}aa})^2 \delta_{ab},\\
&  \sum_{c,d=1}^{\H{N}} d_{\H{N}aa} d_{\H{N}dd} d_{acd} d_{bcd} =
  \frac{N^2-12}{2N} (d_{\H{N}aa})^2 \delta_{ab}, \\
&  \sum_{c=1}^{\H{N}} d_{bcc} (d_{\H{N}bb} + 2 d_{\H{N}cc}) =
  2 \frac{N^2-4}{N} \delta_{\H{N}b},  \\
& 
\begin{aligned}
\sum_{c,d=1}^{\H{N}} (d_{\H{N}cc})^2 d_{acd} d_{bcd} &= \delta_{ab}
  \left( 1 - \frac{8}{N^2} + \delta_{a\H{N}} + \frac{2}{N} \left( \left(
        f_{\H{N}a(a+1)} \right)^2 + \left( f_{\H{N}a(a-1)} \right)^2
    \right) \right.\\
  &
  \left. - \frac{2}{N} (d_{\H{N}aa})^2 + \left( \frac{N}{4} -
      \frac{6}{N} \right) d_{\H{N}aa} d_{\H{N}\H{N}\H{N}}\right),
\end{aligned}\\
&
\begin{aligned}
\sum_{c,d=1}^{\H{N}} d_{\H{N}cc} d_{\H{N}dd} d_{acd} d_{bcd} &=
  \delta_{ab} \left( \frac{1}{2} - \frac{6}{N^2} + \left( \frac{N}{4}
      - \frac{5}{N} \right) (d_{\H{N}aa})^2 - \frac{3}{N} d_{\H{N}aa}
    d_{\H{N}\H{N}\H{N}} \right. \\
  &
  \left. - \left( \frac{N}{4} - \frac{3}{N} \right) \left( \left(
        f_{\H{N}a(a+1)} \right)^2 + \left( f_{\H{N}a(a-1)} \right)^2
    \right) + \left( \frac{3}{2} - \frac{2}{N^2} \right) \delta_{\H{N}a}
  \right),
\end{aligned}\\
&  \sum_{n=1}^{\H{N-1}} \sum_{l=1}^{\H{N}} d_{abl} d_{nnl} = -N
  d_{\H{N}\H{N}\H{N}} d_{\H{N}ab}, \\
&  \sum_{n=1}^{\H{N-1}} \sum_{l=1}^{\H{N}} d_{anl} d_{bnl} = -
  \frac{N}{2} d_{\H{N}\H{N}\H{N}} d_{ab\H{N}} - \frac{3N-2}{N(N-1)}
  \delta_{ab} \sum_{n=1}^{\H{N-1}} \delta_{an} + \frac{N-2}{N-1}
  \delta_{ab} \delta_{a\H{N}} + (N-2) \delta_{ab}. \\
\intertext{An expression related to the last one is,}
&    \sum_{n=1}^{\H{N-1}} \sum_{l=1}^{\H{N}} f_{anl} f_{bnl} =
    \frac{N-2}{2} \sqrt{\frac{2N}{N-1}} d_{ab\H{N}} + \frac{1}{N-1}
    \delta_{ab} \sum_{n=1}^{\H{N-1}} \delta_{an} - \frac{N-2}{N-1}
    \delta_{ab} \delta_{a\H{N}} + (N-2) \delta_{ab}.
\end{flalign}
All of these relations are easily checked numerically for low values
of $N$.

\section*{Acknowledgements}

This work has been partially supported by Conacyt of Mexico through
grant 32598E.

\end{document}